%% file: ms.tex
\documentclass[12pt,preprint]{aastex}

\newcommand{\deltE}{\Delta\kern-1ptE}

\newcommand{\iue}{{\it IUE}}
\newcommand{\hst}{{\it HST}}
\newcommand{\kms}{km~s$^{-1}$}
\newcommand{\gal}{$\alpha$}

\newcommand{\etal}{et al.}
\newcommand{\ecs}{erg cm$^{-2}$ s$^{-1}$}

\begin{document}
\slugcomment{{\bf Astronomical Journal, 2007, (October), in press}}
\title{{\it HST} Observations of Chromospheres in Metal Deficient
Field Giants}

\author{A. K. Dupree, and Timothy Q.  Li}
\affil{Harvard-Smithsonian Center for Astrophysics, \\
Cambridge MA 02138}
\email{adupree@cfa.harvard.edu}

\and
\author{Graeme H. Smith}
\affil{UCO/Lick Observatory, University of California, Santa Cruz, CA 95064}
\email{graeme@ucolick.edu}

\begin{abstract}
HST high resolution spectra of metal-deficient field giants more than 
double the stars in  previous studies, span $\sim$ 3
magnitudes on the red giant branch, and sample an abundance range 
[Fe/H]=$-$1 to $-3$.  These stars, in spite of their age and low 
metallicity, possess  chromospheric fluxes 
of \ion{Mg}{2} ($\lambda$2800) that are within a factor of 4 
of Population~I stars, and give signs of
a dependence on the 
metal abundance at the lowest metallicities.  The \ion{Mg}{2} k-line widths 
depend on luminosity and  correlate with metallicity. 
Line profile asymmetries 
reveal outflows that occur at lower luminosities ($M_V$~=~$-$0.8)  than detected
in Ca K and H-alpha lines in metal-poor giants, 
suggesting mass outflow occurs over a larger 
span of the red giant branch than previously thought, 
and confirming that the \ion{Mg}{2} lines are good
wind diagnostics.  These results do not support a
magnetically
dominated chromosphere, but appear more consistent with some sort
of hydrodynamic, or acoustic heating of the outer atmospheres.

\end{abstract}

\keywords{stars:chromospheres; stars: winds, outflows; stars:
Population II; ultraviolet: stars}

\section{Introduction}

Red giant stars of the Galactic halo have extremes in a number of properties,
such as age, metal abundances, and space motion, that have made them useful
probes of the early evolution of the Galaxy. It is well established that 
the chromospheric activity of stars decreases with increasing age on the main
sequence (e.g., Soderblom 1985; Soderblom \etal\ 1991, Dravins \etal\ 1993) 
In addition, observations show that the fluxes in 
chromospheric and transition region emission lines decrease as stars evolve
into the red giant phase (e.g., Schrijver 1987; Simon \& Drake 1989; 
B\"{o}hm-Vitense 1992). Consequently,
red giants of the Galactic halo present us with the opportunity to study the
extreme limits of another phenomenon, namely stellar activity.

Due to the faintness of even the closest Population II red giants, investigation
of their chromospheres has so far centered around three spectroscopic
features: emission in the wings of the H\gal\ line, in the cores of
the \ion{Ca}{2} H and K lines, and in the cores of the \ion{Mg}{2} {\it h}
and {\it k} lines. Metal-poor giants brighter than 
$\log (L/L_{\odot}) \sim 2.5$ (Cacciari \etal\ 2004) in both the
halo field and globular clusters frequently exhibit emission in one or both 
of the wings of the H\gal\ absorption line (e.g., Cohen 1976; Mallia \& Pagel
1978; Cacciari \& Freeman 1983, Gratton \etal\ 1984; Smith \& Dupree 1988;
Bates \etal\ 1993; Kemp \& Bates 1995).
Lyons \etal\ (1996) conclude that $\approx 80$\% of a total of 68 stars from
five globular clusters with $\log (L/L_{\odot}) > 2.7$ exhibit H\gal\ emission.
The emission profiles are more often than not asymmetric with one wing being
stronger than the other. However the strengths of the emission change with
time. 

Next to the H\gal\ emission features, the most extensively studied
chromospheric diagnostics among Population II giants are the central emission
reversals in \ion{Ca}{2} H and K lines. 
Spectroscopic studies of these emission components 
in both halo field and globular cluster red giants have been reported by 
Dupree \etal\ (1990a),
Smith \etal\ (1992), Dupree \& Smith (1995), and Cacciari \etal\ (2004).
The presence of \ion{Ca}{2} H and K emission is detected further down the
red giant branch of Population II stars than is emission in the wings of
H\gal. The profiles of the Ca emission cores of metal-poor giants are often 
asymmetric, with the sense of the asymmetry changing from blue-dominant to
red-dominant with increasing luminosity on the giant branch.
Chromospheric emission in the H and K lines has also been documented among
metal-poor subdwarfs (Smith \& Churchill 1998), and in the case of at least
one such star, HD 103095, which was included in the Mount Wilson HK survey, 
an activity cycle of period 7.3 yr has been detected (Baliunas \etal\ 1995).

The asymmetries and time variability of the H\gal\ and \ion{Ca}{2} emission
profiles are consistent with the presence of outflows within the chromospheres
of metal-poor stars on the upper giant branch (e.g., Dupree \etal\ 1984; Mauas 
\etal\ 2006). This evidence is supported by the onset of
blueshifts in the H\gal\ at luminosities slightly lower than the onset 
of H\gal\ emission, and blueshifts in  NaD line cores at higher
luminosities  
(e.g., Peterson 1981; Shetrone 1994; Kemp \& Bates 1995;
Shetrone \& Keane 2000; Cacciari \etal\ 2004). Furthermore, wind 
signatures have been found in the $\lambda$10830 \ion{He}{1} line 
of a small number of Population II field giants (Dupree \etal\ 1992;
Smith \etal\ 2004).

The \ion{Mg}{2} chromospheric emission lines, analyzed for cool stars of
approximately solar abundance by Cardini (2005), have been less studied among
metal-poor giants where metallicity effects can be addressed.  
Work based on \iue\ observations of field giants has been 
presented by Dupree \etal\ (1990b), while a \hst\ GHRS spectrum of HD 216143 
was discussed by Smith \& Dupree (1998). The \ion{Mg}{2} emission
in red giants of the globular cluster NGC 6752 have been studied using both
\iue\ and \hst\ GHRS (Dupree \etal\ 1990a, 1994). Despite having 
metallicities of a factor of 100 or more less than solar, the \ion{Mg}{2} 
emission fluxes among the Dupree \etal\ (1990b) sample were in some cases
not substantially less than those of Population I giants and supergiants.
Cuntz \etal\ (1994) found that this relative insensitivity of emission 
flux to metallicity can be accounted for by an acoustic-heating model for
red giant chromospheres.

High-resolution \ion{Mg}{2} \hst\ spectra of nine metal-poor subdwarf stars 
with $-2.2 < {\rm [Fe/H]} < -0.6$ were obtained by Peterson \& Schrijver 
(1997). They found that at ${\rm [Fe/H]} < -1.0$
the emission fluxes were no more than a factor of
three less than that of the solar-like dwarf \gal\ Cen A. Their observations 
suggest that the subdwarf progenitors of metal-poor halo red giants have
significant levels of chromospheric activity, perhaps even comparable to 
quiet regions on the Sun. They suggested that these subdwarfs also have 
acoustically-heated chromospheres. 

The available evidence shows therefore that metal-poor halo stars, despite 
being of great age, do nonetheless have chromospheres with levels of activity
perhaps not very dissimilar from that of the quiet Sun. The strongest of this
evidence comes from the fluxes of the \ion{Mg}{2} {\it h} and {\it k} emission 
lines, which have profiles that show more contrast with the photospheric 
spectrum than either the \ion{Ca}{2} H-K or H\gal\ lines. Given that much of
the \ion{Mg}{2} spectroscopy of metal-poor red giants came from the \iue\
spacecraft, this is an area in which \hst\ archival data can contribute
substantially.

\section{Data}

High resolution spectra of the $\lambda$2800 \ion{Mg}{2} lines in metal-poor 
red giants were taken from  the \hst\ archives for the metal-poor red giants 
listed in Table 1. The position of the stars on a color-magnitude diagram is
shown in Fig. 1, where the dataset spans more than 3 magnitudes for
stars on the red giant branch and asymptotic giant branch, and
abundances
are less than solar by factors of 10 to 1000.  The
spectra for all but three of these stars were obtained during 
\hst\ Cycle 8 as part of a study of the abundances of {\it r}-process heavy 
elements in metal-poor red giants (Proposal 8342, PI J. Cowan). The E230M echelle 
grating was used with {\it HST}-STIS to obtain spectra covering the wavelength 
range $\lambda\lambda$2410-3070 with a resolution of 0.09 \AA\ at the 
$\lambda$2800 \ion{Mg}{2} lines. The STIS spectrum of HD 122563 was 
obtained with a similar setup as part of Program 8111 (PI C. Sneden). Abundance 
analyses based on these spectra are discussed by Cowan \etal\ (2005).
By contrast, GHRS spectra of HD 6833 and HD 216143 were obtained from 
the datasets of Proposals 6010 and 6511 (PIs J. Bookbinder and G. Smith 
respectively). The \ion{Mg}{2} spectrum of HD 216143 from these data has 
been previously discussed by Smith \& Dupree (1998).
Specific details of the individual spectra are given in Table 2.

Spectroscopic metallicities are available in the literature for all of the
stars in Table 1. The [Fe/H] metallicities listed in column 9 come from three 
different sources referred to in column 10, namely
Fulbright (2000), Johnson (2002), and McWilliam \etal\ (1995b), all of whom
employed high-resolution spectra. In the case of the latter two
of these references the quoted value of [Fe/H] is based on analysis of
\ion{Fe}{1} lines. To give a sense of the uncertainties in these metallicities,
the values of [Fe/H] from Cowan \etal\ (2005) are listed in column 11 for
comparison with the other literature values.

The spectra are shown in a stacked presentation in Figures 2 and 3, aligned
according to a photospheric wavelength scale.
The zero-point for each spectrum (marked by broken lines) 
is offset from that of the spectrum below it
by an amount equal to $2 \times 10^{-13}$ \ecs. In addition,
some of the spectra with the lowest flux levels have been further scaled by an 
arbitrary factor.  The expected positions  of the 
interstellar \ion{Mg}{2} {\it h} and {\it k} absorption
features are shown by vertical lines. In many cases these are  shifted
well away from the stellar chromospheric emission profiles as a result of  
large stellar radial velocities.  Thus, the typical central 
reversal of chromospheric emission line profiles becomes apparent in
these objects.  However, for some stars (many of the stars in Figure 2 and 
HD 115444 in Figure 3), the ISM absorption is superimposed
on the chromospheric lines, and in giants such as HD 122563 and HD 165195 
significantly alters the apparent line asymmetry and affects the
observed line fluxes.
 
In most cases the stars exhibit asymmetric {\it h} and {\it k} profiles with 
the sense of the asymmetry being consistent between the two lines. Two notable
exceptions are HD 122956 and HD 6833. In the former case the profiles are
nearly symmetric and the degree of asymmetry is mild. In the case of HD 6833
the {\it k} line is pronouncedly asymmetric ($V/R < 1$) while the {\it h} line 
shows a modest asymmetry in the opposite sense. By contrast, {\it IUE} spectra 
obtained in 1984 showed similar {\it h} and {\it k} profiles (see Figure 2 
of Dupree \etal\ 1990b). Evidently the asymmetry of this star is time 
variable. In the
case of the star $\alpha$ Ori a self-absorbed \ion{Mn}{1} resonance line 
absorbs the short wavelength emission of the \ion{Mg}{2} {\it k} line
(see Figure 5 of Lobel \& Dupree 2000) leading to an apparent asymmetry.
However, this effect is not likely to be the cause of the red asymmetry in
the \hst\ spectrum of the HD 6833 {\it k} line because this line is
narrower than for $\alpha$ Ori so that the \ion{Mn}{1} line lies outside the
emission core.  The k line is formed at higher atmospheric levels than
the h line which could support different velocity fields in a dynamic atmosphere.

\section{Emission Line Fluxes}

The chromospheric fluxes in the \ion{Mg}{2} emission cores were measured
from the spectra by summing directly over the emission profiles between the
flux minima on either side without correcting for any 
photospheric contribution to the line. While this procedure is straightforward, 
it will overestimate the amount of chromospheric emission by including
some photospheric contribution near the line profile minima.
The photospheric \ion{Mg}{2} lines are extremely deep, however, practically 
reaching the zero level near line center, and as such are expected to
contribute a relatively small amount to the derived fluxes. 

To investigate this point, R. Kurucz 
kindly computed synthetic spectra for two of the hottest stars in our sample,
HD 6755 and HD 175305, for which the greatest photospheric fluxes 
would be expected within the wavelength ranges encompassed by the emission
cores. Pure radiative models were constructed using the latest version
of  Kurucz (2005) model atmospheres.  Such models
contain an isothermal atmosphere in hydrostatic equilibrium extended
until the density falls to
negligible values.  The assumption is that no additional heating is 
present to produce a rise in atmospheric temperature.  A  
chromosphere is absent and the synthesized spectrum
represents only the photospheric contribution. The Mg and Ca
abundances 
for these models were enhanced by
0.3 dex over the Fe  values based on the fine analysis by Fulbright (2000).
The atmospheric parameters (Carney \etal\ 2003) adopted for these 
models were $T_{\rm eff} = 5080$K, [Fe/H] = --1.7, $\log g = 2.7$ 
(corresponding to $M_V=1.5$), and
$v_{\rm tot} \sin i = 3.5$ \kms\ for HD 6755, with
$T_{\rm eff}=5050$ K, [Fe/H]= --1.4, $\log g =2.8$
($M_V=1.8$), and $v_{\rm rot}\sin i = 1.5$ \kms\ for HD~175305.
Although we have adopted a small rotational velocity for HD~6755
this star is a spectroscopic binary (Carney \etal\ 2003). However, 
with a period of 1641 days 
it shouldn't be rotating very fast, although there is the uncertainty of
whether the spectrum contains some light from the companion.

The Kurucz photospheric models are overplotted on the observed \ion{Mg}{2} 
spectra in Figures 4 and 5. In both cases, over the wavelength range
bounded by the flux minima on either side of the emission profile,
the net flux in the photospheric \ion{Mg}{2} lines is relatively small 
by comparison with the integrated chromospheric emission.     
Subtracting the contribution of
the synthesized photospheric flux model to the integrated \ion{Mg}{2}
emission amounts to a correction of
less than 12 percent.  For the remainder of the stars, we fit
a local continuum, in the form of a second-order polynomial,
to the minima on each side of
the emission (the k$_1$ and h$_1$ locations).  The difference between
the total flux measured from zero and the continuum-removed flux can amount to
30\%. One star, HD 122563, is an exception where it is a factor
of 2.1 smaller due to the high continuum level (see Figure 2), and this
continuum subtracted flux is used here (Table 3).  
However, this star, and four others have radial velocities 
comparable to that of the local interstellar medium. As a result, 
the observed emission is 
reduced by interstellar \ion{Mg}{2} absorption.  A previous study (Dupree
\etal\ 1990b) of the
effects of interstellar absorption on high latitude metal deficient stars with low
reddening, concluded that the stellar
flux could be increased by $\sim$25\% if interstellar absorption 
were absent.   We have marked these
stars in Table 3, and  the fluxes are displayed in the figures as lower limits.

The {\it k} and {\it h} emission fluxes measured from the \hst\ spectra 
are listed in columns 2 and 3 of Table 3.
The total observed flux $F_{obs}$ for the two lines combined is listed 
in column 4.
These fluxes at Earth were converted to fluxes at the surface of each star
through use of the Barnes \etal\ (1978) relation to calculate the
stellar angular diameter $\phi$ (in units of 10$^{-3}$
arcsec) from the stellar $(V-R)_0$ color. These diameters are listed in 
column 5 of Table 3. A stellar surface flux in units of \ecs\ was then 
calculated from the relation
\begin{equation}
F_{\rm Mg\ II}=
F_{obs}(d/R_{\star})^2 = F_{obs}\times 1.702 \times 10^{17}/\phi^2,
\end{equation}
where $R_{\star}$ is the stellar radius, 
and corrected for reddening according to the $E(B-V)$ values from Table 1 and
$A_{\lambda 2800} = 6.1 E(B-V)$ (Seaton 1979; Cardelli \etal\ 1989).
The values of $F_{\rm Mg\ II}$ are listed in column 6 of Table 3, while column 7
gives their ratio to the solar \ion{Mg}{2} emission flux.

The \ion{Mg}{2} fluxes of the metal-poor giants are much less than that of the
Sun as a star, but the relevant comparison for our purposes is to 
other red giant stars of higher metallicity.
The observed \ion{Mg}{2} surface fluxes measured from the {\it HST} spectra 
are plotted versus $(V-R)_0$ in Figure 6. This figure also includes data
from Dupree \etal\ (1990b) for metal-poor field giants observed 
with {\it IUE} (Table 4), updated using values of $E(B-V)$ and [Fe/H] from
Pilachowski \etal\ (1996), McWilliam \etal\ (1995a,b), Fulbright \etal\ (2000),
Beers \etal\ (2000), and Burris \etal\ (2000). In addition, the fluxes
compiled by Dupree \etal\ for Population I supergiants, Hyades giants, 
and M67 giant stars (Tables 5, 6  and 7) are also shown.
Dupree \etal\ (1990b) found that the surface fluxes of
metal-poor giants in their smaller sample were 
comparable to those of near-solar abundance giants.
A similar effect can be seen in Figure 6 with most of the metal-poor
giants falling close to the Population~I sequence except for stars 
near $(V-R)_0 \sim 0.8$. Four metal-poor
stars in this region have lower limits on their surface
fluxes.  While the correction for interstellar \ion{Mg}{2} 
absorption will increase the stellar flux, it does not appear likely
that the flux will increase by 
0.5 dex, or a factor of 3 or more to bring them up to the Population I 
stars.  Interestingly, the M67 giants, which are
about 0.2 mag redder in $(V-R)_0$ show a similar trend.

Despite the similarity between the \ion{Mg}{2} surface fluxes of metal-poor
and Population~I giants, 
these fluxes do show a correlation with [Fe/H] as
shown in Figure 7. In fact they appear to show less scatter in this figure
at a given [Fe/H] than in Figure 6 at a fixed $(V-R)_0$, suggesting that
the trend seen with [Fe/H] may be more than an artifact of the relation between
red giant color and metallicity.\footnote{In fact, a color-metallicity relation
coupled with the trend seen in Figure 6, would cause a more metal-poor (bluer)
star to have a higher surface flux than a metal-richer giant, the opposite of
what is seen in Figure 7.} Based upon this evidence, the giants which fall 
below the mean sequence in Figure 6 do so as a result of this metallicity
effect. The  most discrepant metal-poor stars in Figure 6 with 
$(V-R)_0 < 0.87$ and \ion{Mg}{2} surface 
fluxes less than $10^5$ \ecs\ are HD 6268, 115444, 122563, 
126587, and 186478. They all have metallicities of ${\rm [Fe/H]} < -2.5$.
Some of the scatter in Figures 6 and 7 might be due to variability. High 
quality {\it IUE} spectra of cool stars have shown a difference up to
a 
factor of 2 in fluxes of the \ion{Mg}{2} lines, 
measured at different times when the expected error of measurement is 
$\leq \pm 20$\% (Brocius \etal\ 1985; Dupree \etal\ 1987).

\section{Line Profiles}

\subsection{Asymmetries}

The asymmetries of the \ion{Mg}{2} {\it k}-line profiles from the {\it HST} 
spectra are shown in Figure~8 as a function of position in the color-magnitude
diagram using the values in Table 1. Stars are plotted with symbols 
designating the $V/R$ parameter, which 
refers to the ratio between the maximum intensities in the violet 
(short-wavelength) and red (long-wavelength) peaks
of the {\it k}-line emission
profile. The 8 stars are plotted whose radial velocities place them away 
from the local interstellar \ion{Mg}{2} absorption, the presence of which 
could affect the asymmetry. The most luminous stars show outflow signatures,
in that the violet emission is less than the red emission. The long 
broken line at $M_V = -1.7$ marks the lower 
extent of the region where asymmetries of $V/R < 1$ and K$_3$ 
absorption reversal blue-shifts dominate in the \ion{Ca}{2} 
K$_2$ emission-line profile as determined from the metal-poor field giant 
spectroscopy of Smith \etal\ (1992) and Dupree \& Smith (1995).  
In the globular cluster NGC 2808, which has a metallicity of [Fe/H] = --1.14
(Carretta \etal\ 2004), Cacciari \etal\ (2004) find a similar limit to
the $V/R < 1$ asymmetry of the Ca II K-line. The \ion{Mg}{2} emission is 
formed higher in the extended stellar atmosphere, and is consequently 
more sensitive to accelerating outflows than the emission lines formed in the
lower chromosphere.  Population I giants show the same signature in 
\ion{Mg}{2} at similar luminosities (Stencel \& Mullan 1980; B\"ohm-Vitense 
1981) indicated by the short-dashed line in Figure~8.  Anthony-Twarog
\& Twarog (1994) evaluated the reddening and derived absolute
magnitudes for these stars from {\it uvby} photometry, and 
their values are used in Figure~9. The
position of our target stars with respect to the Population I objects
remains similar to that in Figure~8. 

\subsection{Wilson-Bappu Emission Line Widths}

Widths of the \ion{Mg}{2} {\it k} lines were measured following the precepts
of Cassatella \etal\ (2001). The observed width is taken to be the full 
width of a fitted Gaussian profile (deleting the central absorption
reversal from the fit), measured at the half maximum value of the
observed line peak. Correction was made for the
instrumental broadening by assuming that the measured width is 
a quadratic sum of the intrinsic stellar width $W_0$
and a resolution element. We omitted three stars
where the emission was very weak and the spectrum noisy
(HD 115444) or the interstellar absorption
seriously compromised the wing of the line (HD 6268 and HD 186478).
The line widths are plotted as a function of $M_V$ in Figure~10 using
the Bond (1980) data, and in Figure~11 for absolute 
magnitudes from Anthony-Twarog \& Twarog (1994).
This is the \ion{Mg}{2} equivalent of the Wilson-Bappu (1957) relation
for metal-poor giants.
Also shown in these figures as a broken line is the mean Wilson-Bappu relation 
for Population I giants as derived by Cassatella \etal\ (2001) for a large
sample of stars for which high-resolution {\it IUE} spectra and 
{\it Hipparcos} parallaxes are available. The Cassatella \etal\ relation 
(their equation 3) is
\begin{equation}
M_V  =(34.56 \pm 0.29) -
 (16.75 \pm 0.14) \log (W_0\phantom{i}{\rm km}\phantom{i}{\rm s}^{-1}).
\end{equation}
The dispersions in line width at
a given $M_V$ among the Cassatella \etal\ (2001) sample are shown by the
shaded regions in Figures~10 and 11, which depict the 50$-$67\% extent in the values of
$\log W_0$ among stars within bins of one-magnitude width. 
The metal-poor red giants have {\it k} emission profiles that are 
narrower than the bulk of the Cassatella \etal\ sample (and in some
cases, narrower than all the stars)  and certainly  more narrow than the mean relation for the sample.

A similar situation was found for the
\ion{Ca}{2} K$_2$ lines of metal-poor red giants by Dupree \& Smith (1995).
By contrast, from their spectroscopy of red giants in NGC 2808, Cacciari
\etal\ (2004) found no evidence for a metallicity dependence of the
Wilson-Bappu effect in \ion{Ca}{2}~K$_2$. The stars in this {\it HST} sample 
all have metallicities 
much lower than NGC 2808 ([Fe/H]=$-$1.14, Carretta \etal\ 2004), so 
that our sample would be more likely to reveal a metallicity dependence to
the Wilson-Bappu effect.  The difference between the observed line
width $\log W_0$ and that given by 
equation (2) is plotted against [Fe/H] in Figure 12 for the metal-poor giants
in the {\it HST} sample. This figure does reveal evidence of a correlation,
with the deviation from the Cassatella \etal\ relation being greater for
lower metallicities.  These figures therefore provide evidence for a
metallicity dependence to the Wilson-Bappu effect for the \ion{Mg}{2} 
{\it k} line. Since the line widths of the metal-poor giants are consistently 
less than predicted by the Cassatella \etal\ (2001) relation for their absolute
magnitude, values of $M_V$ would be underestimated if based on the 
observed \ion{Mg}{2}~{\it k} width.

\section{Discussion}

Results from the \hst\ spectra reported in this paper extend and confirm
those found previously (Dupree \etal\ 1990b) in showing 
that the \ion{Mg}{2} chromospheric
emission fluxes of metal-poor Population II red giants are 
comparable to or within a factor of $\sim 4$ of Population I 
giants and supergiants
of similar $(V-R)_0$ color. This behavior, in spite of [Fe/H] ratios
that are factors of 10 to 1000 
less than the Population I stars and that our sample is  older,
suggests a different chromospheric structure and heating mechanisms from 
Population I stars must be present in these halo red giants.  
The \hst\ data do provide the first 
evidence that the \ion{Mg}{2} emission fluxes correlate with [Fe/H],
at least for [Fe/H]~$< -2.0$ (Figure 7). At higher metallicities this trend
may flatten out although the number of stars in our sample is really too small 
to be definitive on this point.  This metallicity effect may provide an 
explanation for the range in $\log F_{\rm Mg\ II}$ seen among stars with 
$(V-R)_0 \sim 0.7$ to 1.0 in Figure 6. One limitation of the present \hst\
sample is that the color range is fairly restricted, so we cannot conclude very
much about metal-poor giants outside this range, except that the four such
stars in Figure 6 do superimpose on the Population I sequence.

Although the most metal-poor stars in our sample have ${\rm [Mg/H]}< -2$ 
(such stars have ${\rm [Mg/Fe]} \leq 0.4$; Hanson \etal\ 1998), they
have \ion{Mg}{2} emission fluxes no less than about 0.6-0.7 dex that of
Population I giants and supergiants. Peterson \& Schrijver (1997) 
reported similarly that for subdwarfs the \ion{Mg}{2} emission flux did not 
scale strongly with metallicity. They suggested that subdwarfs may have
chromospheres similar to that of the quietest regions of the Sun. The picture
that emerges for metal-poor Population II stars is that their chromospheres
may evolve analogously to those of quiet Population I stars. The similarity in
$F_{\rm Mg\ II}$ between the M67 and the least active metal-poor giants in
Figure 6 may be consistent with this suggestion.  
  
The metal-poor giants in Figure 6 are both older and much more metal-deficient 
than the evolved Population I stars to which they are compared. Magnetic 
processes, direct or indirect, play a role in chromospheric heating of a star
like the Sun (Judge \etal\ 2003), but the contribution of heating 
through dissipation of acoustic waves remains controversial (Cuntz \etal\ 2007; 
Fossom \& Carlsson 2006;
Wedemeyer-B\"ohm \etal\ 2007).  Since the stellar dynamo within a main sequence
star does decline with time it may be that in Population II stars the trends
in chromospheric \ion{Mg}{2} emission are being driven largely by an acoustic
heating of the chromosphere, as argued by Peterson \& Schrijver (1997). Variable
profile asymmetries of H\gal\ in metal deficient giants led Smith \& Dupree (1988)
to identify pulsation as a likely source of heating and momentum deposition.
A  consideration here could involve the energy balance.  If the energy
input were similar in both metal-rich and metal-poor atmospheres, the fact
that radiative cooling is decreased in the metal-poor chromospheres
could lead to a warmer atmosphere and increase the \ion{Mg}{2} emission,
thus compensating for the decreased Mg abundance.

Cuntz \etal\ (1994) have explained the insensitivity of the \ion{Mg}{2}
emission-line fluxes of cool giant stars to metallicity in terms of acoustic 
shock-wave heated chromospheric models. These models account for a 
basal, i.e., non-magnetic, component (e.g., Schrijver
1987; Rutten \etal\ 1991) in the fluxes of the \ion{Mg}{2} and \ion{Ca}{2} 
emission lines of late-type stars (Buchholz \etal\ 1998). Cuntz \etal\
computed the run of physical conditions with height within red giant
chromospheres of various metallicity into which the same acoustic flux is input 
in the form of propagating shock waves. As the metallicity
of the chromosphere is reduced the bulk of the \ion{Mg}{2} {\it k} line
emission tends to be formed deeper at higher mass column densities.
Cuntz \etal\ (1994) computed the physical height in their models
at which the optical depth across the {\it k} line profile (computed
in plane-parallel geometry) was equal to unity
for metallicities ranging from $Z_{\odot}$ to $Z_{\odot}/100$.
Across this resulting range in heights the 
temperature and velocity structure of their shock models did not vary greatly 
with either metallicity or altitude. Thus even though the physical line 
formation depth does vary with metallicity, the line was always formed under 
conditions of similar temperature and
velocity structure. As a result,  the emission flux was found not to vary
greatly with metallicity.

Despite the success of these acoustic models, magnetic heating may still play
a role in maintaining the chromospheres of Population II stars, at least
during some phases of their evolution. The evidence of the subdwarf HD 103095 
([Fe/H] = --1.2, Balachandran \& Carney 1996) suggests that
magnetic activity cycles may be present on the main sequence (Baliunas 
\etal\ 1995), although with a color of $B-V = 0.75$ this
star is much cooler than the main sequence progenitors of the red giants in
Table 1. Main sequence turnoff stars with a metallicity of [Fe/H] = $-$2.3,
[$\alpha$/Fe] = 0.3, and ages of 12$-$14 Gyr have colors of 
$(B-V)_0 \sim 0.325$-0.35 and effective temperatures of 
$\log T_{\rm eff} {\rm (K)} \sim 3.83$ to 3.82 (see Figures 11 and 12 of 
Bergbusch \& VandenBerg 2001), i.e., $T_{\rm eff} = 6760$-6610 K.  These
are similar in temperature to low-mass Population I F2-F5 V stars (Bessell 1979). 
Simon \& Landsman (1991) found that \ion{C}{2} 1335 \AA\ emission among
Population I F dwarfs earlier than spectral type F5 is uncorrelated with 
rotation, supporting the idea that the chromospheres of A and early-F dwarfs 
are heated by the shock dissipation of acoustic waves.
Thus the main sequence progenitors of the metal-poor giants in Table 1
are expected to have had acoustically-heated chromospheres.
When they evolve off the main sequence towards cooler effective temperatures
such stars may transition to chromospheres that have a dynamo contribution
to the heating, as suggested by Simon \& Drake (1989) for Population I giants.
Clump giants in the Hyades clearly give evidence for magnetic activity 
(Baliunas \etal\  1983).  By the time that they evolve beyond the 
luminosity of the horizontal branch, however, it may be that old metal-poor
giants have returned to chromospheres that are mainly heated by
acoustic processes.

Among Population I giants evidence exists for both acoustic and magnetic 
heating of the chromosphere. Observations of Population I G and K giants in 
similar evolutionary states do show evidence for 
correlations between chromospheric emission line 
fluxes, X-ray fluxes, and rotation (Strassmeier \etal\ 1994; Gondoin 2005).
The evolution of X-ray emission with advancing evolution away from the main
sequence depends on stellar mass; for low-mass stars ($M < 1.5$ $M_{\odot}$)
the X-ray activity decreases with evolution away from the main sequence,
while for intermediate-mass giants 
($1.5 \phantom{i} M_{\odot} < M < 3.0 \phantom{i} M_{\odot}$)
the activity at first increases up to a spectral type of around K1, beyond
which it drops (Pizzolato \etal\ 2000; Gondoin 1999, 2005).
During the red giant phase of evolution of Population I stars up to
colors as red as $B-V \sim 1.2$, a dynamo may contribute to chromospheric 
activity, but with advanced evolution to the higher luminosities and cooler 
temperatures of the later K giants the radiative energy losses in the 
\ion{Mg}{2} and other emission
lines can approach the ``basal'' levels observed among red giants by Schrijver
(1987) and Judge \& Stencel (1991), which they attribute to acoustic shock
heating. The evolution of \ion{Ca}{2} H and K emission fluxes among the red
giants of the solar-age open cluster M67 seems consistent with this scenario
(Dupree \etal\ 1999).
The chromospheric evolution of Population II giants may be analogous to those
of Population I, with the modification that the greater ages of the progenitors
of the former stars has reduced the magnetic-heating contribution, so that
the acoustic component is relatively more important.

It will be important to use the next generation of HST instruments to
probe the onset of outflows to fainter magnitudes and access a  
more homogeneous sample of stars represented by globular cluster
members.

\acknowledgements  
We are grateful to Chris Sneden who called our attention to these
spectra from his program and made them available for analysis.  The SIMBAD
reference catalogue produced by the Centre de Donn\'ees astronomiques
de Strasbourg was extremely useful.  The
MAST Archive at STScI provided the HST spectra.
This work was supported in part by NASA.  

{\it Facility:} \facility{HST (GHRS, STIS)}, \facility{IUE}

{}

\clearpage

%FIGURE 1
\begin{figure}
\plotone{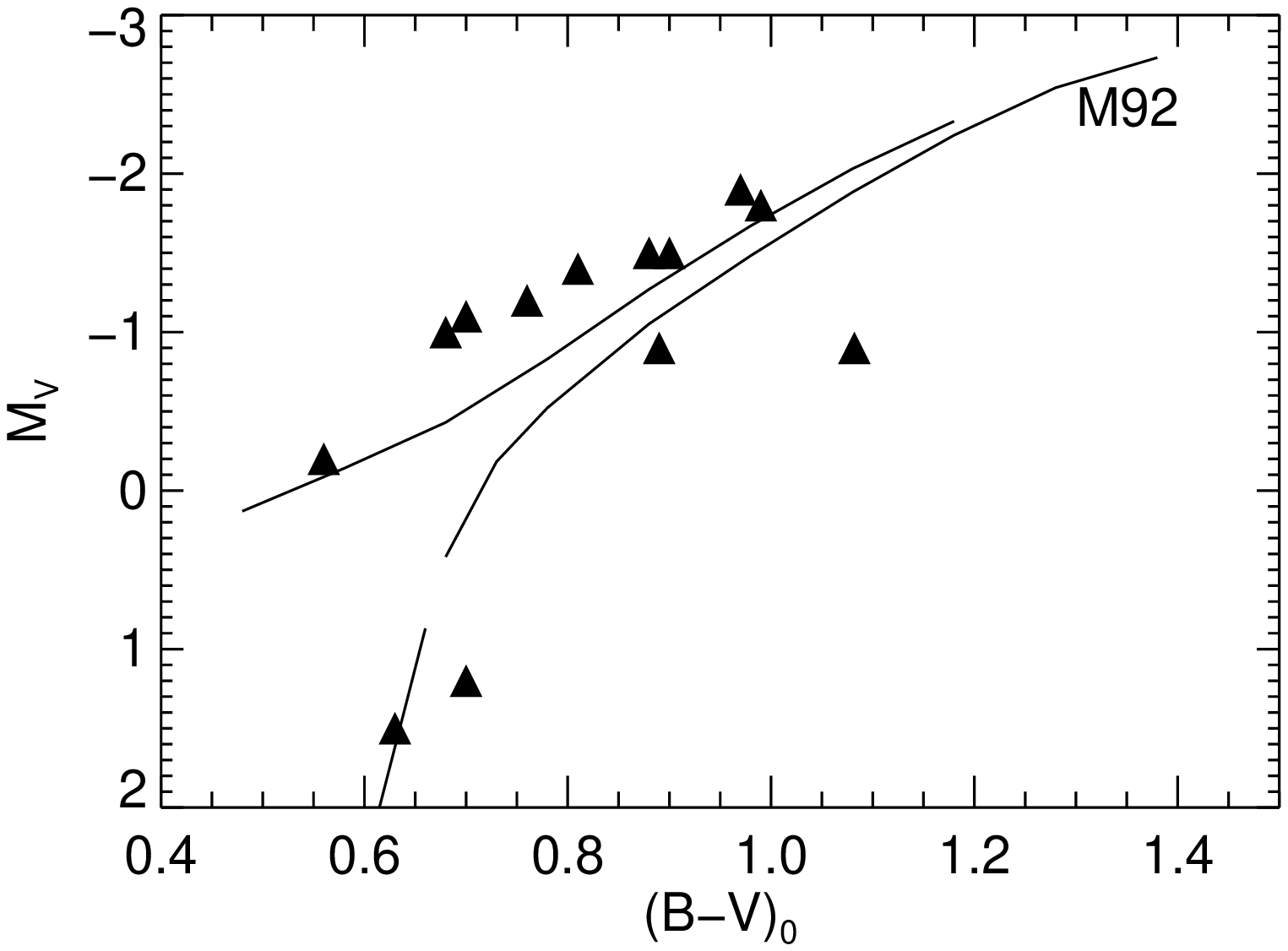}
\caption{Color-magnitude diagram showing the location of the stars 
discussed  in this paper. Solid lines mark the
location of the red giant branch, asymptotic giant branch, and subgiant 
branch in the globular cluster M92 according to Sandage (1970).}
\end{figure}

\clearpage

%FIGURE 2
\begin{figure}
\epsscale{.9}
\plotone{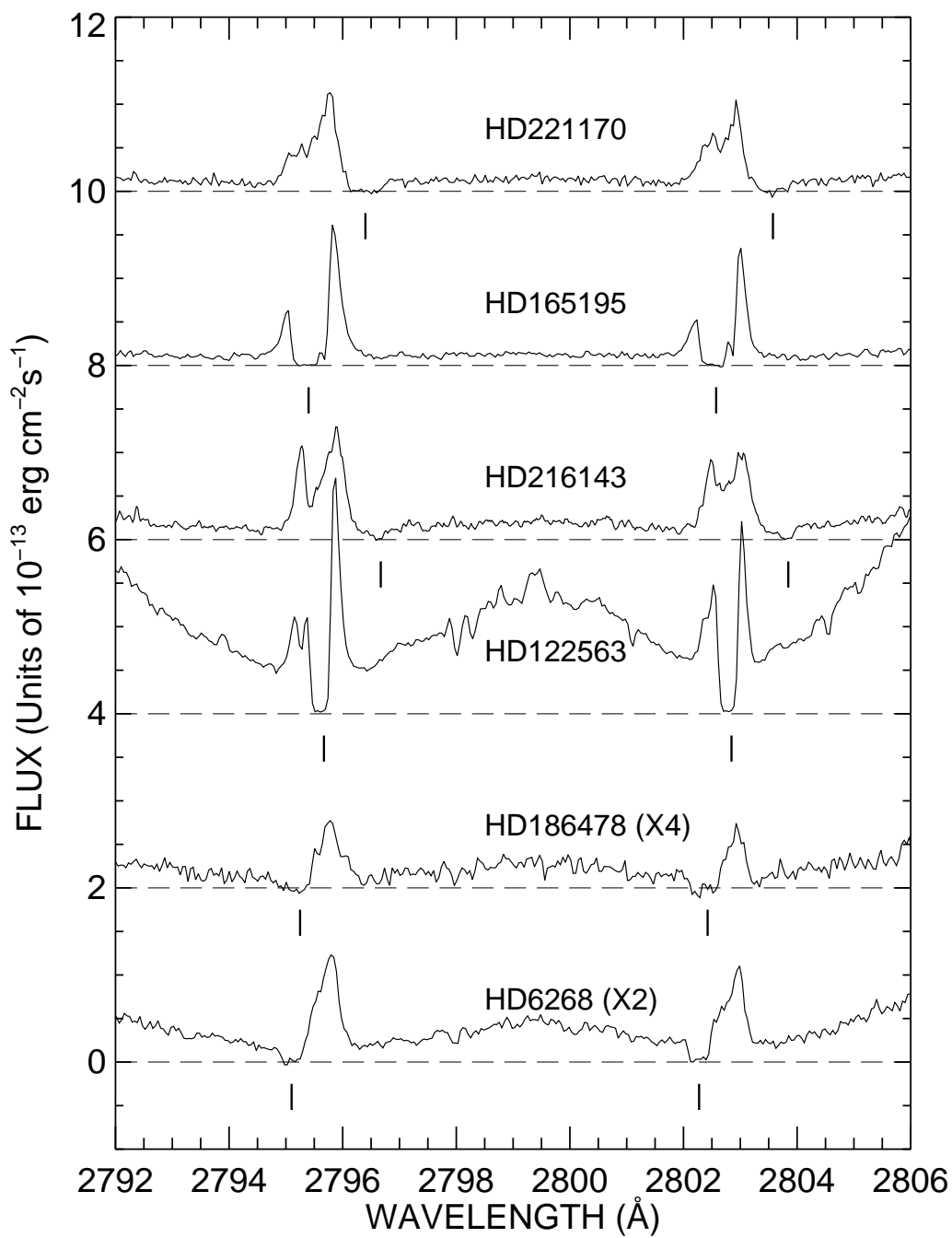}
\epsscale{1}
\caption{\small Chromospheric emission profiles of the 
\ion{Mg}{2} {\it h} and {\it k} lines for stars listed in Table 1.
The spectra are arranged from top to bottom in order of decreasing
absolute visual magnitude.  HD 22170 is the brightest star in the
sample at M$_V$ = $-$1.9 and HD 6268 has M$_V$ = $-$1.2.
The spectra are aligned on a photospheric wavelength scale.  
Vertical lines mark the positions of interstellar medium
absorption. The flux zero-points are offset by $2 \times 10^{-13}$ between 
adjacent spectra, and the spectra of two stars are shown magnified by a
factor of 2 or 4.  The broken line marks the zero level for each spectrum.}
\end{figure}

\clearpage

%FIGURE 3
\begin{figure}
\plotone{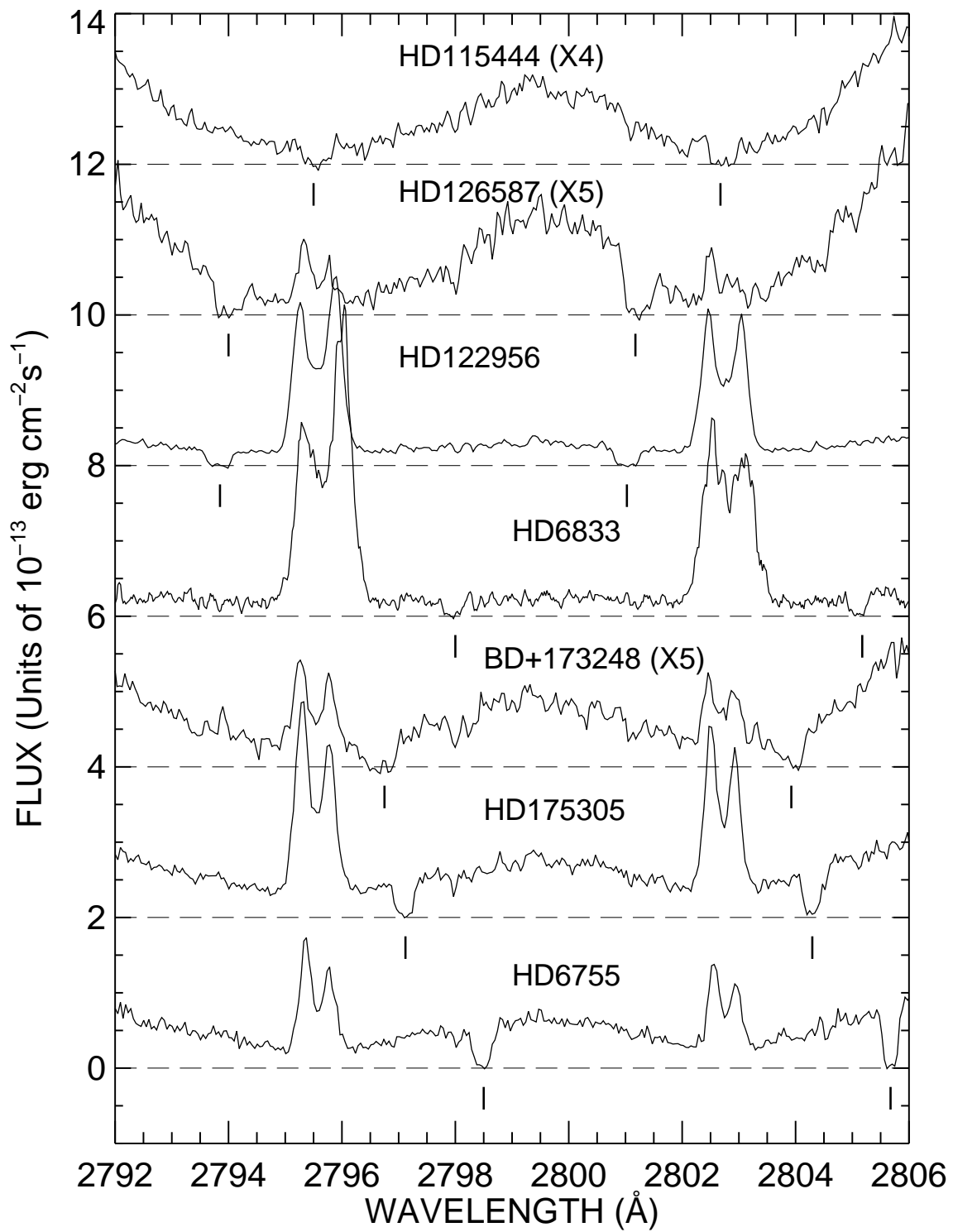}
\caption{Chromospheric emission profiles for the remaining stars
from Table 1, with details the same as Fig. 2.  Here HD 115444 has
M$_V$ = $-$1.1 
and HD 175305 and HD 6755 are the two faintest stars in our sample
at M$_V$ = $+$1.2 and M$_V$ = $+$1.5 respectively.  }
\end{figure}

\clearpage

%FIGURE 4
\begin{figure}
\includegraphics[angle=90,scale=0.6]{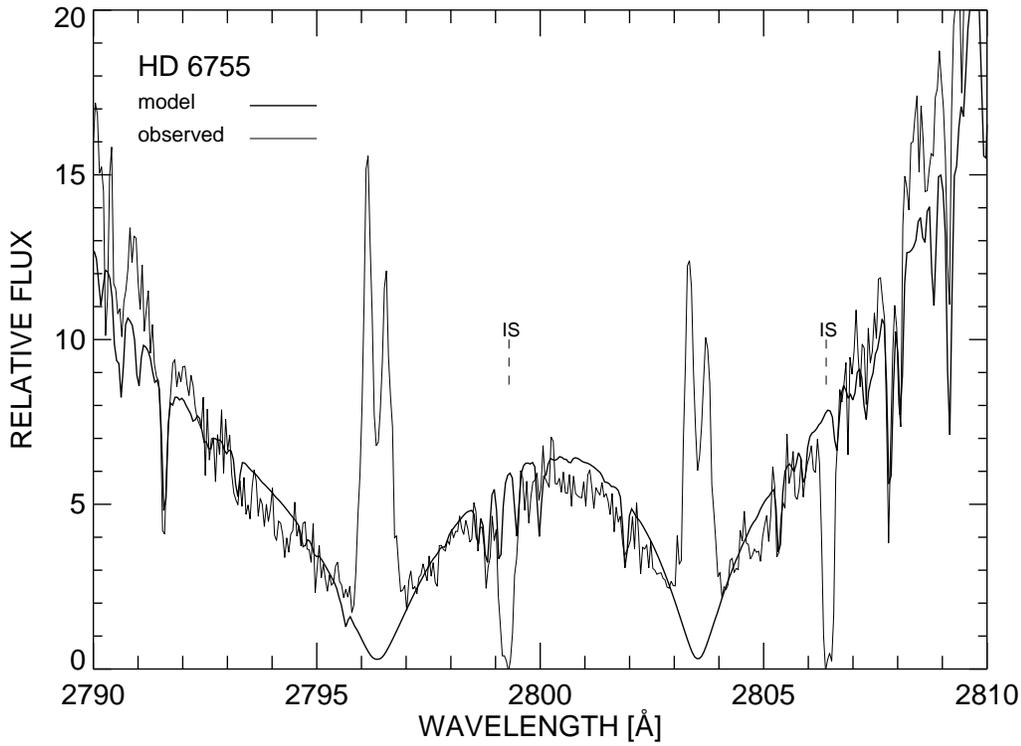}
%\plotone{f4.eps}
\caption{Kurucz model synthesis of a photospheric spectrum for HD 6755 
overplotted on the observed {\it HST STIS} spectrum. The agreement between
the two spectra is quite good outside of the chromospheric emission profiles.  
The interstellar \ion{Mg}{2} lines are marked as IS.}
\end{figure}

\clearpage

%FIGURE 5
\begin{figure}
\includegraphics[angle=90,scale=0.6]{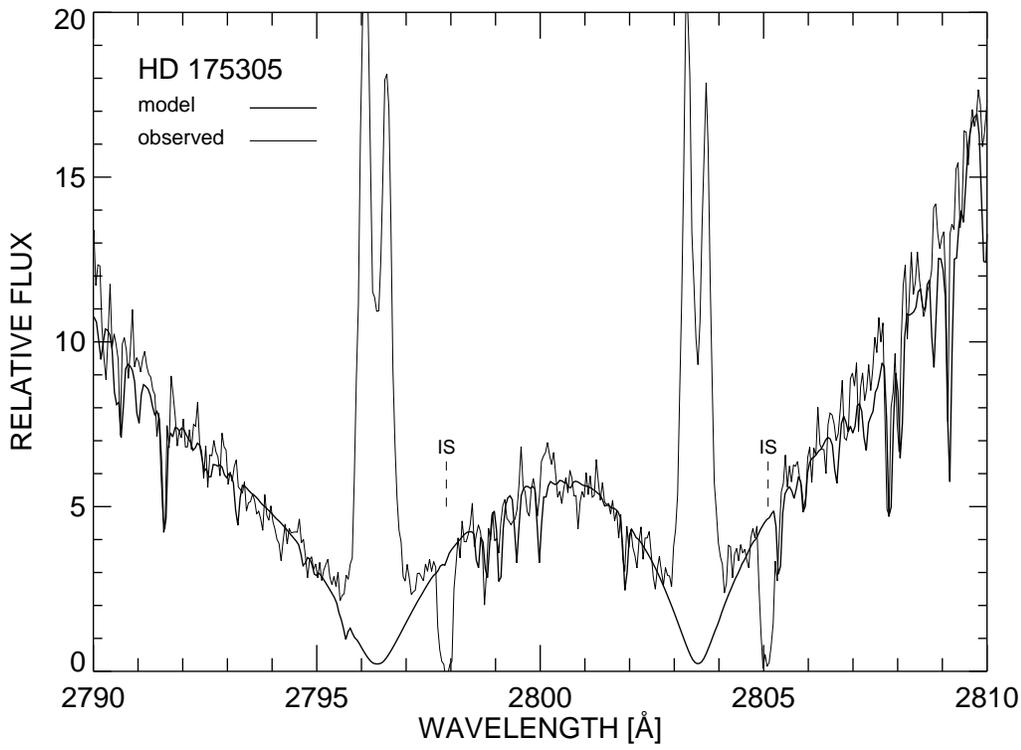}
%\plotone{f5.eps}
\caption{Kurucz model synthesis of a photospheric spectrum for HD 175305 
overplotted on the {\it HST STIS} spectrum.  IS marks the interstellar
\ion{Mg}{2} absorption lines.}
\end{figure}

\clearpage

%FIGURE 6
\begin{figure}
\includegraphics[angle=90,scale=0.8]{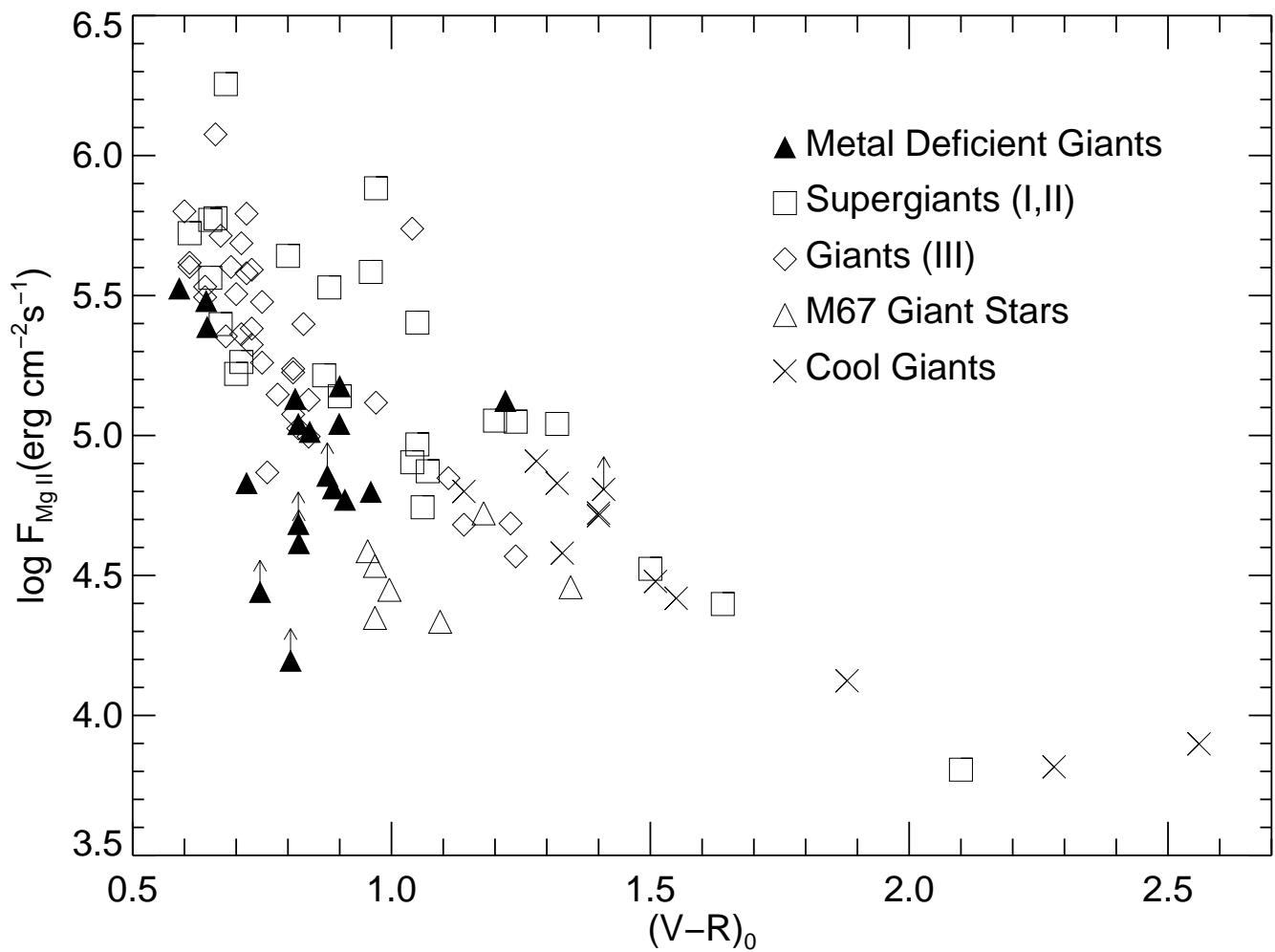}
%\plotone{f6.eps}
\caption{Surface \ion{Mg}{2} fluxes from the {\it HST} sample of this paper
versus $(V-R)_0$ color. Also shown are data for metal
deficient giants reported previously, and Population I field and cluster 
giants, from Dupree \etal\ (1990b). The length of the arrows for lower
limits to the flux represents a 30\% increase in the flux.}
\end{figure}

\clearpage

%FIGURE 7
\begin{figure}
\plotone{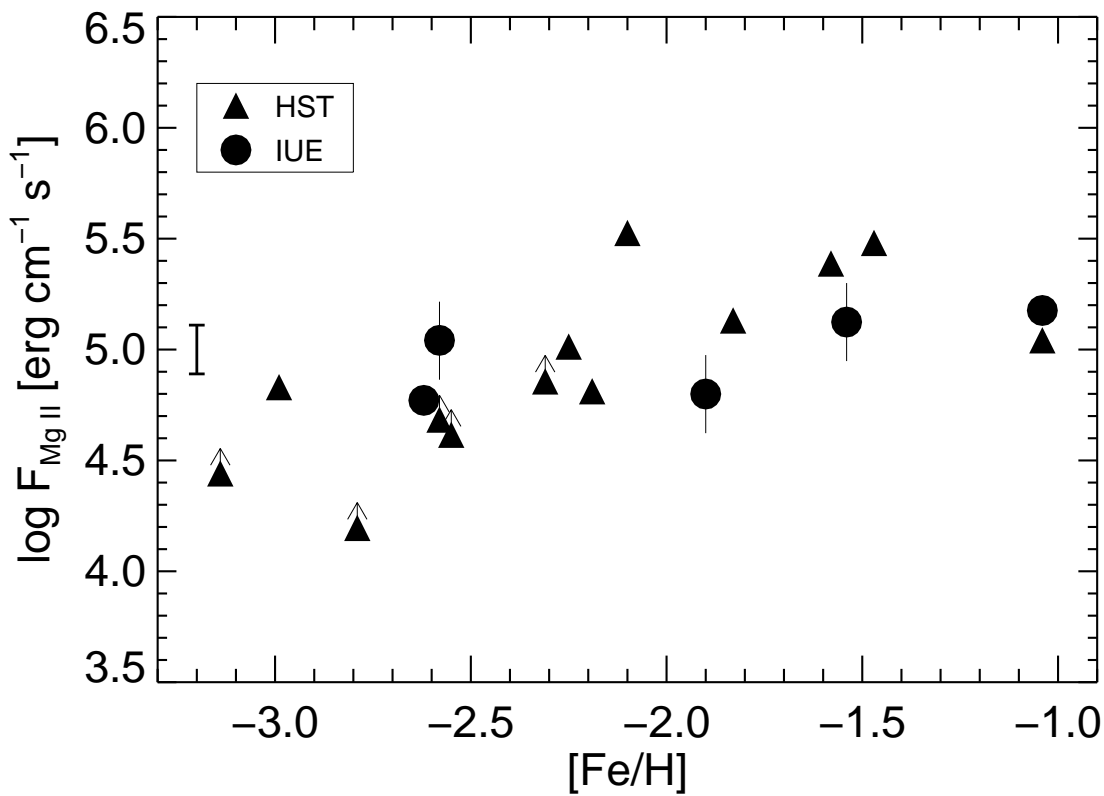}
\caption{Surface \ion{Mg}{2} fluxes of the metal deficient giant stars in 
this sample from {\it HST} and those studied previously with {\it IUE} 
(Dupree \etal\ 1990b) as a function of metallicity.
The error bar shown corresponds to a typical expected error of 
$\pm 30\%$.  Several of the IUE targets had flux errors of $\pm 50\%$ and
those are plotted over the symbol.  Stars whose profiles may be affected by 
interstellar  \ion{Mg}{2} absorption are shown as lower limits, and the 
length of the arrow
represents
a 30\% increase in the flux. }
\end{figure}

\clearpage

%FIGURE 8
\begin{figure}
\plotone{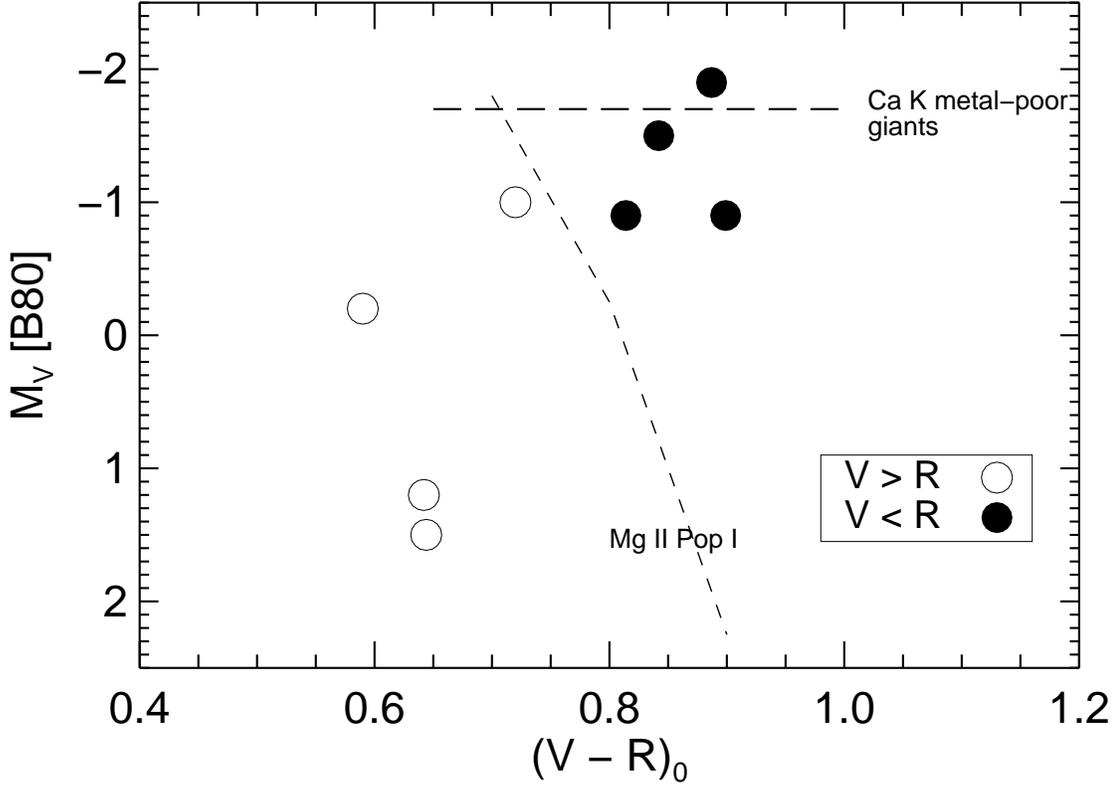}
\caption{Asymmetry of the \ion{Mg}{2} {\it k}-line emission  
derived from the {\it HST} spectra 
as a function of position in the color-magnitude diagram. The
short-dashed line marks the boundary of \ion{Mg}{2} asymmetries in Population
I stars (Stencel \& Mullan 1980; B\"ohm-Vitense 1981); signatures of
outflow occur towards higher (V$-$R) values. 
The long-dashed line marks the lower limit  
of the region where outflow asymmetries ($V/R < 1$) 
and K$_3$-line blueshifts were found in the profile of the 
\ion{Ca}{2} K$_2$ line of metal-poor red giants by 
Smith \etal\ (1992), Dupree \& Smith (1995), and Cacciari \etal\
(2004). 
Values of $M_V$ and reddening are taken from Bond (1980).}
\end{figure}

\clearpage

%FIGURE 9
\begin{figure}
\plotone{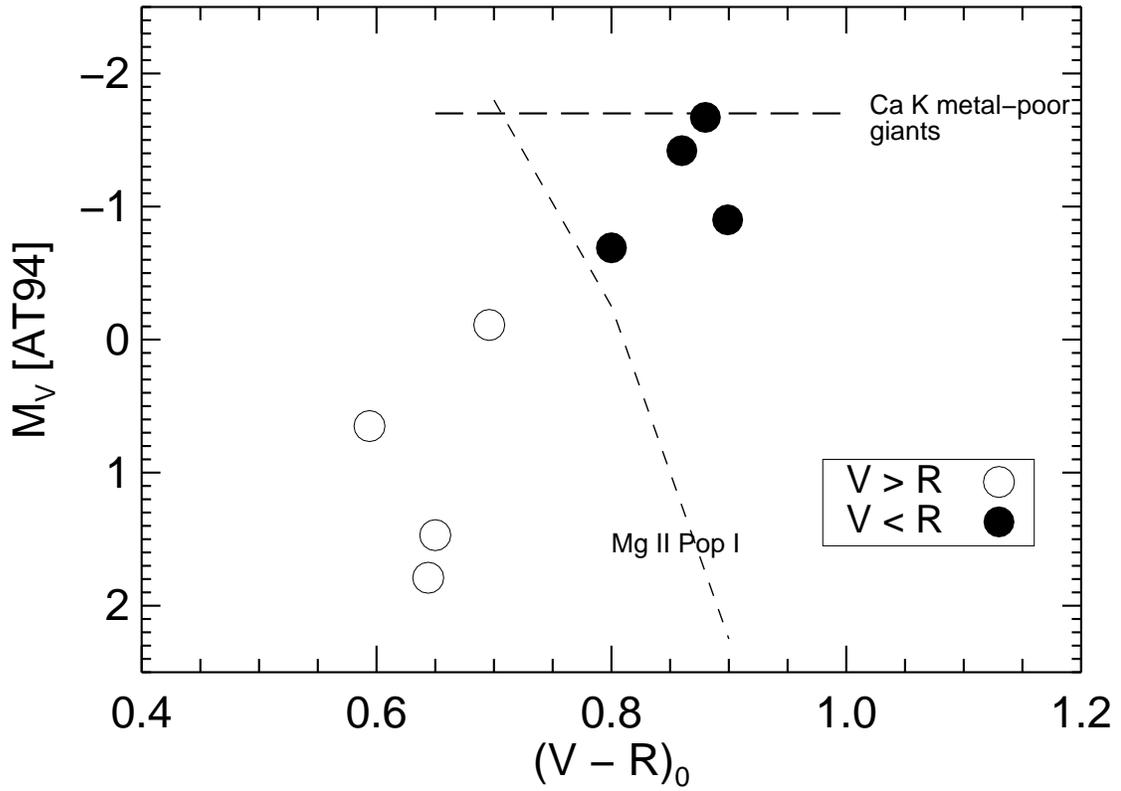}
\caption{Same as Fig. 8, but using the values of $M_V$ and 
reddening from 
Anthony-Twarog \& Twarog (1994) where $E(V-R)=1.1E(b-y)$ 
except for HD 6833 where the Bond (1980) values are assumed.  The
placement of the stars with respect to the asymmetries found in
Population I  stars is similar to that using the Bond (1980)
values.}
\end{figure}
\clearpage

%FIGURE 10
\begin{figure}
\plotone{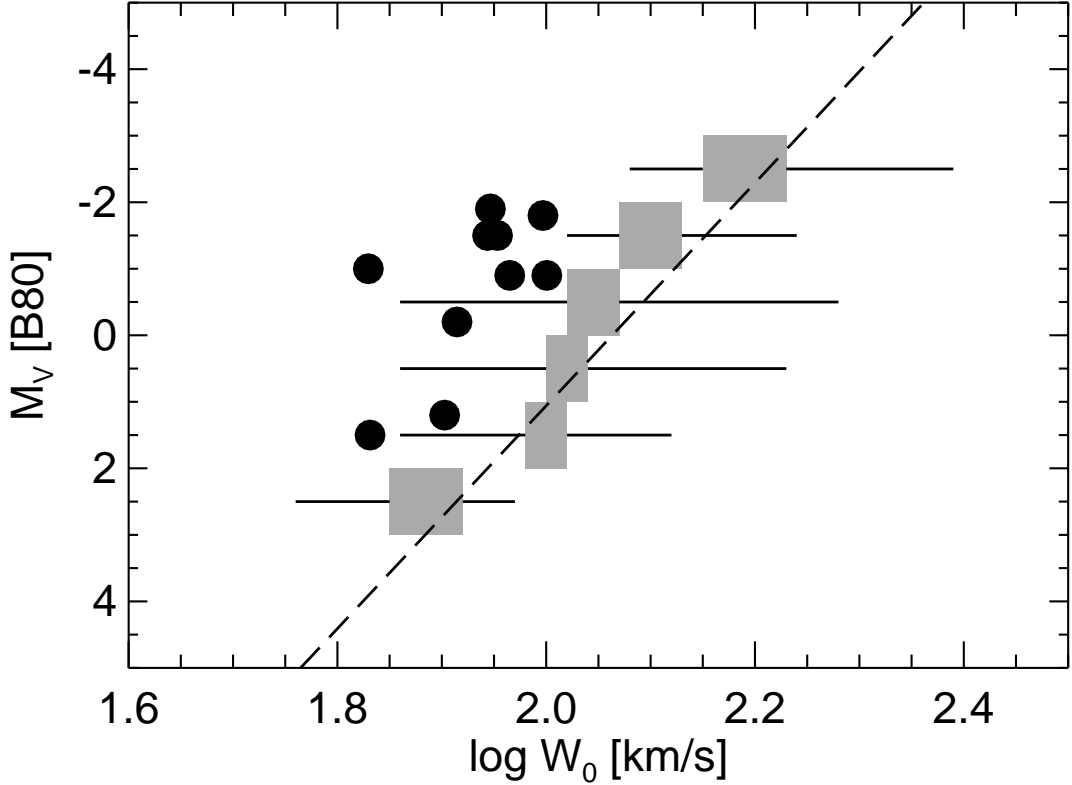}
\caption{Wilson-Bappu width $W_0$ of the \ion{Mg}{2} {\it k} line 
for the metal-poor giants observed by {\it HST} (filled
circles).   The shaded region indicates the
middle 50\% to 67\% of the Cassatella \etal\ sample 
in $\log W_0$ for the stars in one-magnitude $M_V$-bins between 
$M_V$=$\pm$3.  The horizontal lines mark the total extent of the 
measurements of line width, again
for one-magnitude $M_V$-bins. The broken line represents the relation obtained by
Cassatella \etal\ (2001) for a large sample of stars observed
with {\it IUE} and {\it Hipparcos}, including additional brighter and
fainter stars not shown in this figure. The metal-deficient giants in our
sample, here using the values of $M_V$ from Bond (1980)
systematically exhibit smaller values of $\log W_0$.  }
\end{figure}

\clearpage
%FIGURE 11
\begin{figure}
\plotone{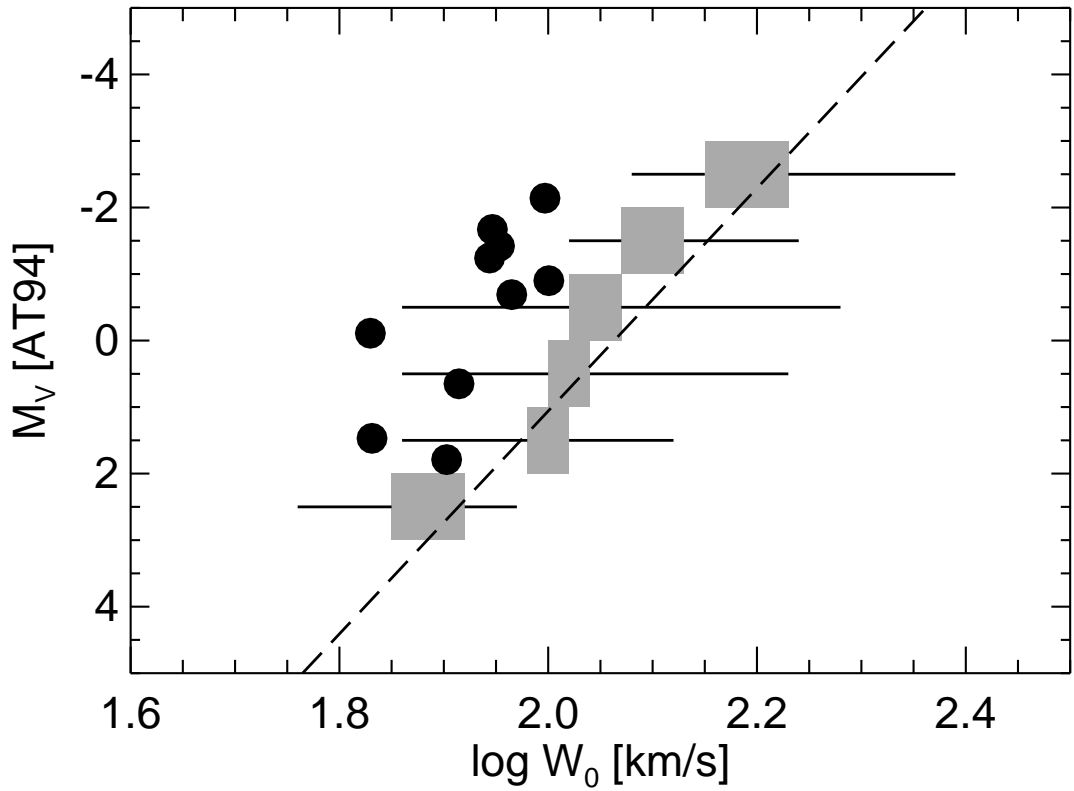}
\caption{Same as Fig. 10 but here using the $M_V$ values
obtained by Anthony-Twarog \& Twarog (1994) except for
HD 6833 where the Bond (1980) values are used.  The line widths
are systematically smaller than found for Population I field
stars.} 

\end{figure}

\clearpage

%FIGURE 12
\begin{figure}

\plotone{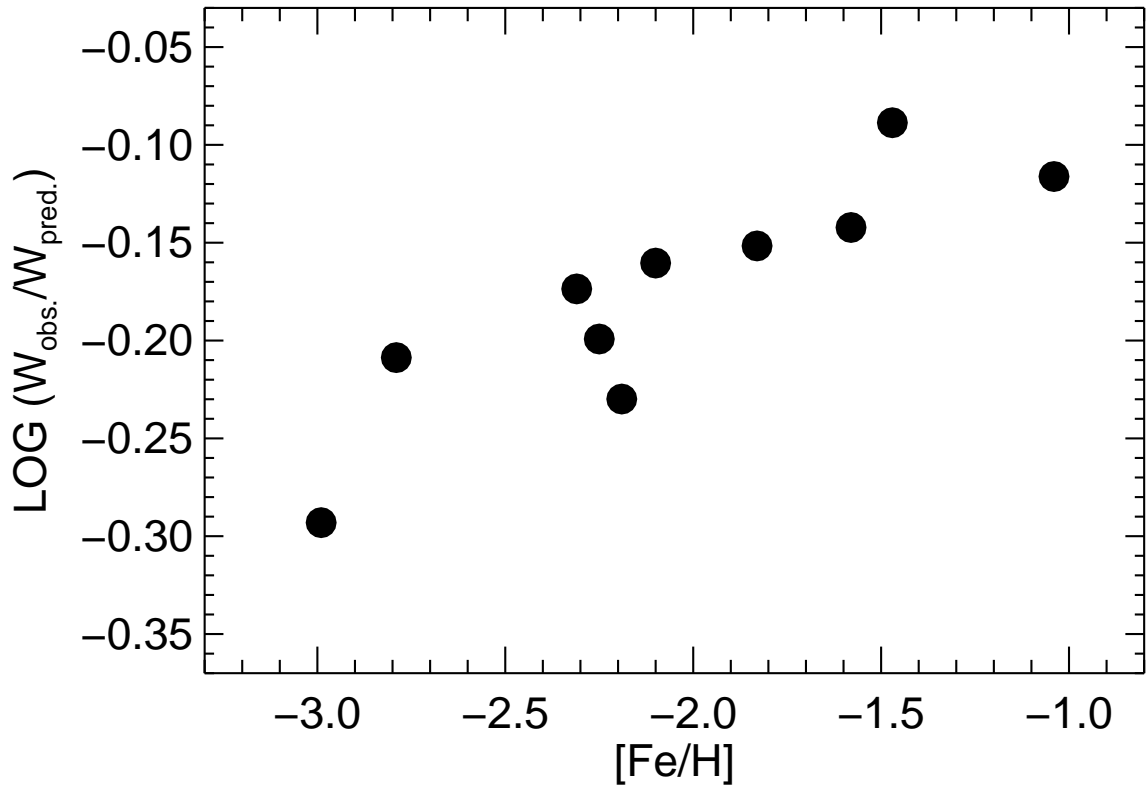}
\caption{Difference between the observed \ion{Mg}{2} 2795 \AA\ line 
width and that predicted by the Cassatella \etal\ (2001) relation 
as a function of [Fe/H] for the metal-poor stars in the {\it HST}
sample
and selecting the Bond (1980) values (Fig. 10). 
The trend here suggests that the deviation is metal dependent.}
\end{figure}

\clearpage

%TABLES
\input{tab1.tex}

\clearpage

\input{tab2.tex}

\clearpage

\input{tab3.tex}

\clearpage

\input{tab4.tex}

\clearpage
\input{tab5.tex}

\clearpage
\input{tab6.tex}

\clearpage
\input{tab7.tex}

\end{document}

%% file: tab1.tex
%\documentclass[12pt,preprint]{aastex}
%\newcommand{\etal}{et al.}
%\begin{document}

\begin{deluxetable}{lcccccccrccc}
\tabletypesize{\scriptsize} 
\tablecolumns{11} 
\tablewidth{0pt} 
\tablecaption{Parameters of Target Stars}
\tablehead{ 
\colhead{Star} & 
\colhead{$V$\tablenotemark{a}} &
\colhead{$(V-R)_J$\tablenotemark{b}}   & 
\colhead{$(B-V)$\tablenotemark{c}} & 
\colhead{$E(B-V)$\tablenotemark{a}} & 
\colhead{$M_V$\tablenotemark{a}} &
\colhead{$V_0$\tablenotemark{d}} & 
\colhead{$(V-R)_0$\tablenotemark{e}} & 
\colhead{Velocity} & 
\colhead{[Fe/H]} &
\colhead{[Fe/H]} & 
\colhead{[Fe/H]} \\
\colhead{}&\colhead{}&\colhead{}&\colhead{}&\colhead{}&\colhead{}&\colhead{}&\colhead{}&
\colhead{(km s$^{-1}$)\tablenotemark{f}}&\colhead{}&\colhead{Ref.\tablenotemark{g}}&\colhead{Ref.\tablenotemark{h}} \\
\colhead{(1)} & \colhead{(2)} & \colhead{(3)} & \colhead{(4)} & \colhead{(5)} & \colhead{(6)} & \colhead{(7)} &
\colhead{(8)} & \colhead{(9)} & \colhead{(10)} & \colhead{(11)} & \colhead{(12)}
}

\startdata
HD 6268\tablenotemark{i}   & 8.11  & \nodata & 0.79 & 0.03 & $-$1.2 & 8.01 & 0.820\tablenotemark{j} & $+$38.4 & $-$2.58 & 1 & $-$2.42 \\
HD 6755   & 7.73  & 0.676 & 0.67 & 0.04 & $+$1.5 & 7.60 & 0.644 & $-$319.2 &  $-$1.58 & 2 & $-$1.68 \\
HD 6833   & 6.75  & 0.947 & 1.14 & 0.06 & $-$0.9 & 6.55 & 0.899 & $-$245.0 &  $-$1.04 & 2 & \nodata \\
HD 115444 & 8.98  & 0.746 & 0.70 & 0.00 & $-$1.1 & 8.98 & 0.746 &  $-$28.0 &  $-$3.15 & 3 & $-$2.90 \\
HD 122563 & 6.21  & 0.805 & 0.90 & 0.00 & $-$1.5 & 6.21 & 0.805 &  $-$23.2 &  $-$2.79 & 2 & $-$2.72 \\
HD 122956 & 7.22  & 0.846 & 0.93 & 0.04 & $-$0.9 & 7.09 & 0.814 & $+$166.0 &  $-$1.83 & 2 & $-$1.95 \\
HD 126587 & 9.12  & 0.76\ \tablenotemark{k} & 0.73 & 0.05 & $-$1.0 &
8.96 & 0.72\ \tablenotemark{k} & $+$149.0 &  $-$2.99 & 3 & $-$2.93 \\
BD $+$17 3248\tablenotemark{i} & 9.40 & 0.638 & 0.62 & 0.06 & $-$0.2 & 9.20 & 0.590 & $-$146.0 &  $-$2.11 & 2 & $-$2.08 \\
HD 165195 & 7.34  & 1.076 & 1.24 & 0.25 & $-$1.8 & 6.52 & 0.876 &   $-$0.2 &  $-$2.32 & 3 & \nodata \\
HD 175305 & 7.20  & 0.666 & 0.73 & 0.03 & $+$1.2 & 7.10 & 0.642 & $-$181.0 &  $-$1.47 & 2 & $-$1.48 \\
HD 186478 & 9.16  & 0.893 & 0.90 & 0.09 & $-$1.4 & 8.86 & 0.821 &  $+$31.0 &  $-$2.58 & 1 & $-2.56$ \\
HD 216143 & 7.82  & 0.874 & 0.94 & 0.04 & $-$1.5 & 7.69 & 0.842 & $-$116.0\tablenotemark{a} &  $-$2.25 & 2 & \nodata \\
HD 221170 & 7.68  & 0.927 & 1.02 & 0.05 & $-$1.9 & 7.52 & 0.887 & $-$119.0 &  $-$2.19 & 2 & $-$2.08 \\

\enddata

\tablenotetext{a}{Bond (1980).}
\tablenotetext{b}{Color index from Stone (1983) on the Johnson system.}
\tablenotetext{c}{SIMBAD database.}
\tablenotetext{d}{$V$ magnitude corrected for interstellar extinction assuming $A_V = 3.3E(B-V)$.}
\tablenotetext{e}{Assuming $E(V-R)/E(B-V)=0.8$.} 
\tablenotetext{f}{SIMBAD database, with the exception of HD 216143,
which was acquired from Bond (1980).}
\tablenotetext{g}{References for [Fe/H] in column 10: (1) McWilliam \etal\ 1995b; (2)
Fulbright 2000; (3) Johnson 2002.}
\tablenotetext{h}{[Fe I/H] abundances from Cowan \etal\ (2005).}
\tablenotetext{i}{Bond notes as AGB star.}
\tablenotetext{j}{Dupree \etal\ (1990b).}
\tablenotetext{k}{Carney (1980).}

\end{deluxetable} 

%\end{document}
%% 
%% End of file `table1.tex'. 

%% file: tab2.tex
%\documentclass[12pt,preprint]{aastex}
%\begin{document}

\begin{deluxetable}{lrrrrrrrrr}
\tabletypesize{\scriptsize}
\tablecolumns{6}
\tablewidth{0pt}
\tablecaption{HST Observations}

\tablehead{
\colhead{Star} & 
\colhead{Dataset} &
\colhead{Date Observed} &
\colhead{Exposure} &
\colhead{Data Type}&
\colhead{Mg II K} \\
\colhead{}& \colhead{}& \colhead{}& \colhead{(s)}& \colhead{}&
\colhead{Asymmetry\tablenotemark{a}}
}

\startdata
HD 6268   & O5F603010  & 2001 Jan 31 & 1870 & STIS/E230M & \nodata \\
          & O5F603020  & 2001 Jan 31 & 2923 & STIS/E230M & \\
HD 6755   & O5F654010  & 2000 Mar 27 & 2040 & STIS/E230M & $V>R$\\
HD 6833   & Z3GA0106T  & 1996 Oct 08 & 577 & GHRS/G270M & $V<R$\\
HD 115444 & O5F601010  & 1999 Jul 16 & 1882 & STIS/E230M & \nodata\\
          & O5F601020  & 1999 Jul 16 &11908 & STIS/E230M & \\
          & O5F602010  & 1999 Jul 02 & 1882 & STIS/E230M & \\
HD 122563 & O5EL01010  & 1999 Jul 29 & 1767 & STIS/E230M & \nodata\\
          & O5EL01020  & 1999 Jul 29 & 2887 & STIS/E230M & \\
          & O5EL01030  & 1999 Jul 29 & 2868 & STIS/E230M & \\
          & O5EL01040  & 1999 Jul 29 & 2868 & STIS/E230M & \\
HD 122956 & O5F605010  & 1999 Jul 31 & 1858 & STIS/E230M & $V<R$\\
          & O5F605020  & 1999 Jul 31 & 2899 & STIS/E230M & \\
HD 126587 & O5F606010  & 2000 Jul 04 & 1811 & STIS/E230M & $V>R$ \\
          & O5F606020  & 2000 Jul 04 &11624 & STIS/E230M & \\
HD 165195 & O5F608010  & 1999 Oct 16 & 1850 & STIS/E230M & \nodata\\
          & O5F608020  & 1999 Oct 16 & 2899 & STIS/E230M & \\
HD 175305 & O5F609010  & 1999 Aug 22 & 2195 & STIS/E230M & $V>R$\\
HD 186478 & O5F610010  & 1999 Jun 03 & 1802 & STIS/E230M & \nodata\\
          & O5F610020  & 1999 Jun 04 &11588 & STIS/E230M & \\
          & O5F611010  & 1999 Aug 21 & 1802 & STIS/E230M & \\
HD 216143 & Z3EI0105T  & 1996 Aug 05 & 1184 & GHRS/G270M &$V<R$ \\
          & Z3EI0107P  & 1996 Aug 05 & 1523 & GHRS/G270M & \\
          & Z3EI0108T  & 1996 Aug 05 & 870 & GHRS/G270M & \\
HD 221170 & O5F662010  & 2000 Jul 13 & 1894 & STIS/E230M & $V<R$\\
BD $+$17 3248 & O5F607010 & 1999 Oct 15 & 1802 & STIS/E230M & $V>R$ \\
             & O5F607020 & 1999 Oct 15 &  8691 & STIS/E230M & 

\enddata

\tablenotetext{a}{$V$ and $R$ refer to the short-wavelength and long-wavelength sides of the
emission core respectively. Asymmetry in the K-line (2795\AA) emission could not be
assessed in several objects where interstellar absorption affects the line profile.}
\end{deluxetable} 

%\end{document}

%% 
%% End of file `table2.tex'. 

%% file: tab3.tex
%\documentclass[12pt,preprint]{aastex}
%\begin{document}
\begin{deluxetable}{lrrrrrrr}

%\tabletypesize{\scriptsize} 
\tablecolumns{8} 
\tablewidth{0pt} 
\tablecaption{Mg II Fluxes From Metal$-$Deficient Field Giants}

\tablehead{ 
\colhead{}&
\multicolumn{3}{c}{Mg II Observed Flux\tablenotemark{a}} & 
\colhead{} & \colhead{} & \colhead{} &\colhead{}\\
\colhead{Star} & 
\colhead{$\lambda$2795} &
\colhead{$\lambda$2802}   & 
\colhead{Total} & 
\colhead{$\phi$\tablenotemark{b}} &
\colhead{F$_{*}$\tablenotemark{c}} & 
\colhead{F$_{*}$/F$_{\odot}$$^d$}&
\colhead{Log W$_0$\tablenotemark{e}}
}
\startdata
HD 6268   & 3.11e$-$14\tablenotemark{f} & 2.48e$-$14\tablenotemark{f}& 5.59e$-$14\tablenotemark{f} &
0.482& 4.83e+4& 0.039&\nodata\\
HD 6755   & 9.50e$-$14 &7.53e$-$14 & 1.70e$-$13& 0.385&2.44e+5&0.195&1.83 \\
HD 6833   & 3.03e$-$13 & 2.18e$-$13 & 5.21e$-$13& 1.062 &1.10e+5&0.088&2.00 \\
HD 115444 & 5.66e$-$15\tablenotemark{f} & 5.60e$-$15\tablenotemark{f} & 1.13e$-$14\tablenotemark{f}& 0.264&2.76e+4&0.022&\nodata\\
HD 122563 & 6.91e$-$14\tablenotemark{f} & 3.87e$-$14\tablenotemark{f} & 1.08e$-$13\tablenotemark{f}&1.082 & 1.57e+4&0.012&1.94\\
HD 122956 & 1.89e$-$13 & 1.51e$-$13 & 3.40e$-$13&0.732&1.35e+5&0.108&1.97 \\
HD 126587 & 1.04e$-$14 & 8.54e$-$15 & 1.90e$-$14&0.252&6.77e+4&0.054&1.83\\
BD $+$17 3248 & 1.91e$-$14 & 1.51e$-$14 & 3.42e$-$14&0.156&3.36e+5&0.268 &1.91 \\
HD 165195 & 6.35e$-$14\tablenotemark{f} & 4.95e$-$14\tablenotemark{f}& 1.13e$-$13\tablenotemark{f}&1.044 &7.19e+4&0.057&2.00 \\
HD 175305 & 1.90e$-$13 & 1.57e$-$13 & 3.48e$-$13& 0.481&3.04e+5&0.242&1.90 \\
HD 186478 & 9.41e$-$15\tablenotemark{f} & 6.15e$-$15\tablenotemark{f} & 1.56e$-$14\tablenotemark{f}&
0.326& 4.13e+4&0.033 &\nodata\\
HD 216143 & 8.81e$-$14 & 7.43e$-$14 & 1.62e$-$13 & 0.579&1.03e+5&0.082 &1.95\\
HD 221170 & 6.81e$-$14 & 6.05e$-$14 & 1.29e$-$13 &0.670& 6.48e+4&0.052&1.95
\enddata 
\tablenotetext{a}{Units of erg cm$^{-2}$ s$^{-1}$ observed
at Earth. The notation 3.11e$-$14 denotes 3.11 $\times$10$^{-14}$.}
\tablenotetext{b}{Stellar angular diameter (in units of 10$^{-3}$
arcsec) from the Barnes, Evans, \& Moffett (1978) relation based on ($V-R$).}
\tablenotetext{c}{Stellar surface flux 
in units of erg cm$^{-2}$ s$^{-1}$ where 
$F_{\star}=F_{obs}(d/R_{\star})^2 = F_{obs}\times 1.702 \times
10^{17}/\phi^2$.
$F_{\star}$ has been corrected
for reddening (Seaton 1979; Cardelli \etal\ 1989) where $A_{\lambda
2800} =6.1 E(B-V)$. 
Reddenings are listed in Table 1.}
\tablenotetext{d}{Solar flux in Mg II lines taken as 1.25 $\times$
10$^6$ erg cm$^{-2}$ s$^{-1}$.}
\tablenotetext{e}{The corrected full width at half power of the Mg II 2795\AA\ 
line (km s$^{-1}$).}
\tablenotetext{f}{The stellar radial velocity is $\pm$ 50\kms\ or less
and so there is potential for absorption by interstellar Mg II.  In
some case absorption is visible in the line
profile.  This means the observed Mg II fluxes  are lower limits to
the stellar fluxes.}

\end{deluxetable} 
%\end{document}

%% 
%% End of file `table.tex'. 

%% file: tab4.tex
% \documentclass[12pt,preprint]{aastex}
% \newcommand{\etal}{et al.}
% \begin{document}

\begin{deluxetable}{lrlccccccl}
\tabletypesize{\scriptsize} 
\tablecolumns{9 } 
\tablewidth{0pt} 
\tablecaption{Mg II Fluxes (IUE) from Metal-Deficient Giants}
\tablehead{ 
\colhead{HD} &
\colhead{$V$\tablenotemark{a}} &
\colhead{$E(B-V)$\tablenotemark{b}}&
\colhead{$(V-R)_0$\tablenotemark{c}}   & 
\colhead{[Fe/H]} &
\colhead{$\phi$\tablenotemark{c}} & 
\colhead{Mg II Obs.\tablenotemark{d}}  &
\colhead{log$_{10}$F$_\star$\tablenotemark{e}} &
\colhead{Refs.}\\ 
\colhead{}&\colhead{}&\colhead{}&\colhead{}&
\colhead{}&\colhead{(10$^{-3}$ arcs)}&\colhead{(erg cm$^{-2}$ s$^{-1}$)}&
\colhead{(erg cm$^{-2}$ s$^{-1}$)}
}

\startdata
6268& 8.11&0.03&0.82&$-$2.58&0.483 &1.1e-13&5.02&1  \\
6833&6.75&0.06&0.90&$-$1.04&1.062&5.5e-13&5.09&2 \\
84903&8.06&0.13&0.91&$-$2.60&0.648&1.2e-13&5.00&3 \\
135148&9.49&0.04&0.96&$-$1.90&0.311&9.9e-14&4.80&4 \\
232078&8.62&0.5:&1.22&$-$1.54&1.296&8.7e-14&5.17&4 \\
\enddata

\tablenotetext{a}{Bond 1980. }
\tablenotetext{b}{Bond 1980, 
except for HD232078 where $E(B-V)=0.5:$ is taken, between
values of 0.38 (Beers \etal\ 2000) and 0.6 (Gonzalez \& Wallerstein 1998).}
\tablenotetext{c}{Stone 1983; value for HD 84903 from Carney 1980.}

\tablenotetext{d}{Total flux in Mg II h and k emission cores observed
at Earth with IUE, Dupree \etal\ 1990b.}
\tablenotetext{e}{Unreddened flux from stellar surface using the
relations $E_{B-V}=0.355 E_{V-K}$, $A_{\lambda 2800}=6.1E_{B-V}$
(Seaton 1979). }

\tablerefs{(1) McWilliam \etal\ 1995a;
(2) Fulbright 2000;
(3) Beers \etal\ 2000;
(4) Burris \etal\ 2000.}

\end{deluxetable} 
%\end{document}
%% 
%% End of file `table6.tex'. 

%% file: tab5.tex
%\documentclass[12pt,preprint]{aastex}
%\newcommand{\etal}{et al.}
%\begin{document}

\begin{deluxetable}{lrlrcrccc}
\tabletypesize{\scriptsize} 
\tablecolumns{9 } 
\tablewidth{0pt} 
\tablecaption{Mg II Fluxes from Giant and Supergiant Stars}
\tablehead{ 
\colhead{Star} &
\colhead{HD}&
\colhead{Sp. Type}&
\colhead{$V$\tablenotemark{a}} &
\colhead{$(V-R)$\tablenotemark{a}}   & 
\colhead{$\phi$\tablenotemark{b}} & 
\colhead{Mg II Obs.\tablenotemark{c}}  &
\colhead{log$_{10}$F$_\star$\tablenotemark{d}} & 
\colhead{Notes}\\
\colhead{}&\colhead{}&\colhead{}&\colhead{}&
\colhead{}&\colhead{(10$^{-3}$ arcs)}&\colhead{(erg cm$^{-2}$ s$^{-1}$)}&
\colhead{(erg cm$^{-2}$ s$^{-1}$)}&\colhead{}
}

\startdata
$\beta$ Aqr& 204867&G0 Ib& 2.87 &0.61&3.11\phantom{0000}&3.00e-11&5.72&e\\
$\alpha$ Aqr&209750& G2 Ib&2.93& 0.66 & 3.44\phantom{0000} &4.15e-11&5.78&e\\
$\beta$ Dra& 159181&G2 Ib-IIa& 2.78&0.68&3.87\phantom{0000}&1.58e-10&6.26&f\\
$\xi$ Pup& 63700& G3 Ib&3.35&0.88&4.49\phantom{0000}&4.0\phantom{0}e-11&5.53&g\\
9 Peg& 206859& G5 Ib& 4.31&0.80&2.57\phantom{0000}&1.70e-11&5.64&f\\
$\epsilon$ Gem&48329&G8 Ib& 2.98& 0.96& 6.00\phantom{0000}&8.12e-11&5.58&f\\
56 Peg&218356&K0 Ibp&4.77&0.97&2.67\phantom{0000}&3.2\phantom{0}e-11&5.88&g\\
$\epsilon$ Peg& 206778&K2 Ib&2.39&1.05&8.98\phantom{0000}&1.2\phantom{0}e-10&5.40&g\\
$\lambda$ Vel& 78647&K5 Ib&2.21&1.24&12.91\phantom{0000}&1.1\phantom{0}e-10&5.05&f\\
$\xi$ Cyg& 200905&K5 Ib& 3.70&1.20&6.13\phantom{0000}&2.5\phantom{0}e-11&5.05&g\\
$\sigma$ CMa& 52877&K7v Ib&3.43&1.32&8.28\phantom{0000}&4.4\phantom{0}e-11:&5.04&g\\
$\alpha$ Ori&39801&M2 Iab&0.37&1.64&54.3\phantom{0}\phantom{0000}&4.3\phantom{0}e-10&4.40&h\\
\\
$\epsilon$ Leo& 84441&G1 II&2.98&0.65&3.27\phantom{0000}&3.7\phantom{0}e-11&5.77&g\\
$\beta$ Lep& 36079&G5 II& 2.84&0.65&3.49\phantom{0000}&2.62e-11&5.56&f\\
$\delta$ Col&44762&G7 II&3.85&0.67&2.31\phantom{0000}&1.4\phantom{0}e-11&5.40&i\\
$\zeta$ Cyg&202109&G8 II&3.20 & 0.70&3.36\phantom{0000}&1.10e-11&5.22&f\\
$\zeta$ Hya&76294&G8 II-III&3.10&0.71&3.61\phantom{0000}&1.40e-11&5.26&f\\
$\theta$ Lyr&180809&K0 II&4.37&0.87&2.77\phantom{0000}&7.40e-12&5.21&f\\
$\alpha$ Hya&81797&K2 II&1.97&1.04&10.7\phantom{0}\phantom{0000}&5.40e-11&4.90&g\\
$\gamma$ Aql&186791&K3 II&2.72&1.07&8.0\phantom{0}\phantom{0000}&1.4\phantom{0}e-11&4.87&j\\
$\iota$ Aur&31398&K3 II&2.69&1.06&7.9\phantom{00000}&1.0\phantom{0}e-11&4.74&j\\
$\theta$ Her&163770&K3 II&3.87&0.90&3.7\phantom{00000}&5.6\phantom{0}e-12&5.14&j\\
$\alpha$ TrA&150798&K4 II&1.9\phantom{0}&1.05&11.5\phantom{00000}&3.6\phantom{0}e-11&4.97&j\\
$\beta$ Peg& 217906&M2 IIb&2.42&1.50&17.2\phantom{00000}&5.8\phantom{0}e-11:&4.52&g\\
$\alpha$ Her& 156014&M5 II&3.06&2.1:&31.0\phantom{00000}&3.6\phantom{0}e-11:&3.81:&g\\
\\
$\epsilon$ Hya& 74874&G0 III&3.38&0.60&2.40\phantom{0000}&2.14e-11&5.80&e\\
$o$ UMa&71369&G5 III&3.36&0.69&3.04\phantom{0000}&2.2\phantom{0}e-11&5.60&i\\
$\mu$ Vel&93497&G5 III&2.69&0.66&3.84\phantom{0000}&1.03e-10&6.08&f\\
$\eta$ Psc&9270&G7 IIIa&3.62&0.72&2.90\phantom{0000}&1.9\phantom{0}e-11&5.58&i\\
$\beta$ Crv&109379&G7 IIIa&2.64&0.61&3.46\phantom{0000}&2.91e-11&5.62&f\\
$\beta$ Her&148856&G7 IIIa&2.77&0.64&3.51\phantom{0000}&2.26e-11&5.49&f\\
$\iota$ Cnc& 74739&G7.5 IIIa& 4.02&0.75&2.61\phantom{0000}&1.2\phantom{0}e-11&5.48&i\\
$\delta$ Crt& 98430&G8 III-IV&3.56&0.83&3.79\phantom{0000}&2.1\phantom{0}e-11&5.40&i\\
$\eta$ Dra&148387&G8 III&2.74&0.61&3.30\phantom{0000}&2.56e-11&5.60&f\\
$\mu$ Peg&216131&G8 III&3.48&0.68&2.81\phantom{0000}&1.05e-11&5.35&f\\
$\epsilon$ Vir&113226&G8 IIIb&2.84&0.64&3.40\phantom{0000}&2.31e-11&5.53&f\\
$\eta$ Her&150997&G8 IIIb&3.50&0.67&2.71\phantom{0000}&2.23e-11&5.71&f\\
$\delta$ Dra&180711&G9 III&3.07&0.70&3.57\phantom{0000}&2.4\phantom{0}e-11&5.51&i\\
\\
$\alpha$ Phe&2261&K0 III&2.40&0.81&6.28\phantom{0000}&4.00e-11&5.24&f\\
$\alpha$ Cas&3712&K0 IIIa&2.23&0.78&6.44\phantom{0000}&3.40e-11&5.15&f\\
$\nu$ Oph&163917&K0 IIIa&3.34&0.71&3.23\phantom{0000}&1.4\phantom{0}e-11&5.36&i\\
$\alpha$ UMa&95689&K0 IIIa&1.79&0.81&8.31\phantom{0000}&6.80e-11&5.23&f\\
$\theta$ Cen&123139&K0 IIIb&2.06&0.76&6.62\phantom{0000}&1.90e-11&4.87&f\\
$\beta$ Gem&62509&K0 IIIb&1.14&0.75&9.86\phantom{0000}&1.04e-10&5.26&k\\
$\beta$ Cet& 4128&K1 III&2.02&0.72&6.09\phantom{0000}&1.35e-10&5.79&f\\
$\alpha$ Boo&124897&K1+ IIIB&$-$0.05&0.97&24.56\phantom{0000}&4.66e-10&5.12&f\\
$\beta$ Oph&161096&K2 III& 2.77&0.82&5.37\phantom{0000}&1.80e-11&5.03&f\\
$\kappa$ Oph&153210&K2 III& 3.20&0.84&4.54\phantom{0000}&1.20e-11&5.00&f\\
$\alpha$ Ari&12929&K2 IIIab-b&2.00&0.84&7.89\phantom{0000}&4.90e-11&5.13&f\\
$\alpha$ Ser&140573&K2 IIIb&2.64&0.81&5.62\phantom{0000}&2.21e-11&5.08&f\\
$\alpha$ Tuc&211416&K3 III&2.85&1.04:&7.16\phantom{0000}&1.65e-10&5.74&f\\
$\beta$ UMi&131873&K4 III&2.08&1.11&11.32\phantom{0000}&5.30e-11&4.85&f\\
$\gamma$ Dra& 164058&K5 III&2.22&1.14&11.09\phantom{0000}&3.47e-11&4.68&f\\
$\alpha$ Tau&29139&K5 III&0.86&1.23&23.69\phantom{0000}&1.60e-10&4.69&f\\
$\beta$ And&6860&M0 III&2.05&1.24&13.90\phantom{0000}&4.20e-11&4.57&e\\
\enddata

\tablenotetext{a}{From Johnson \etal\ 1966.}
\tablenotetext{b}{Stellar angular diameter obtaained from $V, V-R$
(assumed unreddened), and Barnes-Evans-Moffett relationship (1978).}
\tablenotetext{c}{Total flux in Mg II h and k emission cores observed at Earth.}
\tablenotetext{d}{Flux from stellar surface.}
\tablenotetext{e}{Hartmann \etal\ 1982.}
\tablenotetext{f}{Simon \etal\ 1982.}
\tablenotetext{g}{Stencel \etal\ 1980.}
\tablenotetext{h}{Dupree \etal\ 1987. The Mg II line in $\alpha$ Ori
is variable with a total amplitude of about a factor of 2.}
\tablenotetext{i}{Simon \& Drake 1989.}
\tablenotetext{j}{Hartmann \etal\ 1985. Observed flux in the
Mg II {\it h}-line only.  Flux at stellar surface in Mg h has
been multipled by
a factor of 2 in order to approximate the contribution of the Mg II k
line in the Mg II doublet.}
\tablenotetext{k}{Baliunas \etal\ 1983.}

\end{deluxetable} 

% \end{document}
%% 
%% End of file `table1.tex'. 

%% file: tab6.tex
% \documentclass[12pt,preprint]{aastex}
% \newcommand{\etal}{{\it et al.}}
% \begin{document}

\begin{deluxetable}{lrlcccccc}
\tabletypesize{\scriptsize} 
\tablecolumns{8 } 
\tablewidth{0pt} 
\tablecaption{Mg II Fluxes from Cool Giant Stars}
\tablehead{ 
\colhead{Star} &
\colhead{HD}&
\colhead{Sp. Type\tablenotemark{a}}&
\colhead{$V$} &
\colhead{$(V-R)_0$\tablenotemark{b}}   & 
\colhead{$\phi$\tablenotemark{c}} & 
\colhead{Mg II Obs.\tablenotemark{d}}  &
\colhead{log$_{10}$F$_\star$\tablenotemark{e}} \\ 
\colhead{}&\colhead{}&\colhead{}&\colhead{}&
\colhead{}&\colhead{(10$^{-3}$ arcs)}&\colhead{(erg cm$^{-2}$ s$^{-1}$)}&
\colhead{(erg cm$^{-2}$ s$^{-1}$)}
}

\startdata
87 Leo&99998&K4.5 III&4.76&1.14&3.71&5.1\phantom{0}e-12&4.80\\
74 Gem& 61338&M0.0 III&5.05&1.28&2.97&4.20e-12&4.91\\
HIP 21365&29051&M1.1 III&7.1\phantom{0}&1.32&3.01&3.4\phantom{0}e-13&4.83\\
$\pi$ Leo&86663&M1.7 III&4.70&1.40&4.88&5.70e-12&4.72\\
$\lambda$ Aqr&216386&M2.0 III&3.79&1.40&8.21&1.00e-11&4.71\\
82 Vir&119149&M2.1 III&5.01&1.33&4.34&4.20e-12&4.58\\
$\upsilon$ Cap&196777&M2.1 III&5.17&1.41&4.72&$>$5.00e-12&$>$4.81\\
WW Psc&5820&M2.4 III&6.11&1.55&3.16&1.54e-12&4.42\\
$\psi$ Vir&112142&M2.7 III&4.80&1.51&5.85&3.60e-12&4.48\\
XZ Psc&224062&M4.6 III&5.61&1.88&6.30&3.12e-12&4.12\\
HIP 91781&172816&M5.2 III&6.27&2.28&9.03&1.18e-12&3.82\\
RZ Ari&18191&M5.9 III&5.91&2.56&10.2\phantom{0}&4.8\phantom{0}e-12&3.90\\
\enddata

\tablenotetext{a}{Ridgway \etal\ 1980. }
\tablenotetext{b}{Measured by Barnes \etal\ (1978) or derived from
$(V-K)_0$ values (Steiman-Cameron \etal\  1985) using relation between
$(V-K)_0$ and $(V-R)_0$ tabulated by Johnson \etal\ 1966.}
\tablenotetext{c}{Occultation angular diameter (Ridgway \etal\ 1980). }

\tablenotetext{d}{Total flux in Mg II h and k emission cores observed
at Earth (Steiman-Cameron \etal\ 1985).}
\tablenotetext{e}{Unreddened flux from stellar surface using the
relations $E_{B-V}=0.355 E_{V-K}$, $A_{\lambda 2800}=6.1E_{B-V}$
(Seaton 1979). }

\end{deluxetable} 

%\end{document}
%% 
%% End of file `table6.tex'. 

%% file: tab7.tex
 %\documentclass[12pt,preprint]{aastex}
%\newcommand{\etal}{et al.}
%\begin{document}

\begin{deluxetable}{lrlccrccc}
\tabletypesize{\scriptsize} 
\tablecolumns{8 } 
\tablewidth{0pt} 
\tablecaption{Mg II Fluxes from Cluster Giant Stars}
\tablehead{ 
\colhead{Star} &
\colhead{Other Designation}&
\colhead{Sp. Type}&
\colhead{$V$\tablenotemark{a}} &
\colhead{$(V-R)_0$\tablenotemark{b}}   & 
\colhead{$\phi$\tablenotemark{c}} & 
\colhead{Mg II Obs.\tablenotemark{d}}  &
\colhead{log$_{10}$F$_\star$\tablenotemark{e}} \\ 
\colhead{}&\colhead{}&\colhead{}&\colhead{}&
\colhead{}&\colhead{(10$^{-3}$ arcs)}&\colhead{(erg cm$^{-2}$ s$^{-1}$)}&
\colhead{(erg cm$^{-2}$ s$^{-1}$)}
}

\startdata
\cutinhead{Hyades (C 0424+157)}
$\epsilon$ Tau&HD 28305&G9.5 III&3.54& 0.73&3.10\phantom{0000}&1.19e-11&5.32\\
$\delta$ Tau& HD 27697&K0 III&3.76&0.73&2.80\phantom{0000}&1.11e-11&5.38\\
$\theta$$'$ Tau&HD 28307&K0 IIIb&3.83&0.71&2.58\phantom{0000}&1.90e-11&5.69\\
$\gamma$ Tau&HD 27371&K0- IIIab&3.65&0.73&2.95\phantom{0000}&2.00e-11&5.59\\
\cutinhead{M 67 (NGC 2682)\tablenotemark{f}} 
T-856&BD +12 1917&K2 III:&9.84&0.95&0.269\phantom{000}&1.3\phantom{0}e-14&4.59\\
T-829&BD +12 1913&K2 III&9.53&0.97&0.317\phantom{000}&1.6\phantom{0}e-14&4.53\\
T-626&BD +12 1924&K2 III&9.37&1.09&0.411\phantom{000}&1.7\phantom{0}e-14&4.33\\
F-170&BD +12 1926&K3 III&9.69&0.97&0.294\phantom{000}&9.0\phantom{0}e-15&4.35\\
F-108&TYC 814-2331-1&K4 III&9.72&1.00&0.302\phantom{000}&1.2\phantom{0}e-14&4.45\\
IV-202&BD +12 1919&K5 III&8.86&1.18&0.588\phantom{000}&8.5\phantom{0}e-14&4.72\\
S-1553&BD +11 1934&M0 III&8.74&1.35&0.797\phantom{000}&8.5\phantom{0}e-14&4.46\\
\enddata

\tablenotetext{a}{Hyades: Johnson \etal\ 1966. M67: Janes \&
Smith 1984.}
\tablenotetext{b}{Unreddened colors on Johnson system. Hyades values from Johnson \etal\ (1966) where
E(B$-$V)=0.  M67 values derived from (V$-$R) observed on Kron-Cousins
system and dereddened assuming $E(V-R)/E(B-V)=0.69$ for K5 giants
(Janes \& Heasley 1988), $E(B-V)=0.041$ (Taylor 2007), and the
transformation
between Kron-Cousins and Johnson $V-R$ colors was made using the
equation proposed by Cousins (1976), viz.:
$(V-R)_J=1.40(V-R)_{KC}+0.028$ for $(V-R)_{KC} < 1$. }
\tablenotetext{c}{Stellar angular diameter obtained from $V, (V-R)_0$
and Barnes-Evans-Moffett relationship (1978).}
\tablenotetext{d}{Total flux in Mg II h and k emission cores observed
at Earth. Hyades: Baliunas \etal\ 1983. M67: Smith \& Janes 1988.}
\tablenotetext{e}{Unreddened flux from stellar surface. Stellar
surface flux of M67 stars corrected for reddening by assuming
$A_{\lambda 2800}= 6.1 E(B-V)$ from Seaton (1979) , where $E(B-V)=
0.041$ (Taylor 2007). Hyades: $E(B-V)=0$. }
\tablenotetext{f}{Stars identified in the following catalogues:
T and IV designation: Murray \& Clements (1968); F designation:
Fagerholm stars identified in Johnson and Sandage (1955);
S designation: Sanders (1977).}

\end{deluxetable} 

%\end{document}
%% 
%% End of file `table5.tex'. 

%% file: ms.bbl
\begin{thebibliography}{}


\bibitem[]{}Anthony-Twarog, B. J., \& Twarog, B. A. 1994, \aj, 107,
1577 [AT94]

\bibitem[]{587}Balachandran, S. C., \& Carney, B. W. 1996, \aj, 111, 946

\bibitem[]{589}Baliunas, S. L., Hartmann, L., \& Dupree, A. K. 1983, \apj, 
271, 672

\bibitem[]{592}Baliunas, S. L., \etal\ 1995, \apj, 438, 269

\bibitem[]{594}Barnes, T. G., Evans, D. S., \& Moffett, T. J. 1978, \mnras, 183, 
  285

\bibitem[]{597}Bates, B., Kemp, S. N., \& Montgomery, A. S. 1993, \aap, 97, 937

\bibitem[]{599}Beers, T. C., Chiba, M., Yoshii, Y., Platais, I., Hanson, R. B., 
Fuchs, B., \& Rossi, S. 2000, AJ, 119, 2866

\bibitem[]{602}Bergbusch, P. A., \& VandenBerg, D. A. 2001, \apj, 556, 322

\bibitem[]{604}Bessell, M. S. 1979, \pasp, 91, 589

\bibitem[]{606}B\"{o}hm-Vitense, E. 1981, \apj, 244, 504

\bibitem[]{608}B\"{o}hm-Vitense, E. 1992, \aj, 103, 608

\bibitem[]{610}Bond, H. E. 1980, \apjs, 44, 517 [B80]

\bibitem[]{612}Brosius, J. W., Mullan, D. J., \& Stencel, R. E. 1985,
\apj, 288, 310

\bibitem[]{615}Buchholz, B., Ulmschneider, P., \& Cuntz, M. 1998, \apj, 494, 700

\bibitem[]{617}Burris, D. L., Pilachowski, C. A., Armandroff, T. E.,
  Sneden, C., Cowan, J. J., \& Roe, H.  2000, \apj, 544, 302

\bibitem[]{620}Cacciari, C., \& Freeman, K. C. 1983, \apj, 268, 185

\bibitem[]{622}Cacciari, C., Bragaglia, A., Rossetti, E., Fusi Pecci, F.,
  Mulas, G., Carretta, E., Gratton, R. G., Momany, Y., \& Pasquini, L.
  2004, \aap, 413, 343

\bibitem[]{626}Cardelli, J. A., Clayton, G. C., \& Mathis, J. S. 1989,
  \apj, 345, 245

\bibitem[]{629}Cardini, D. 2005, \aap, 430, 303

\bibitem[]{631}Carney, B. W. 1980, \aj, 85, 38

\bibitem[]{633}Carney, B. W., Latham, D. W., Stefanik, R. P., Laird,
J. B., \& Morse, J. A. 2003, AJ, 125, 293

\bibitem[]{636}Carretta, E., Bragaglia, A., \& Cacciari, C. 2004, \apj, 610, L25 

\bibitem[]{638}Cassatella, A., Altamore, A., Badiali, M., \& Cardini, D.
 2001, \aap, 374, 1085

\bibitem[]{641}Cohen, J. G. 1976, \apj, 203, L127

\bibitem[]{}Cousins, A. W. J. 1976, Mem. R. A. S., 81, 25

\bibitem[]{643}Cowan, J. J., Sneden, C., Beers, T. C., Lawler, J. E., Simmerer, 
 J., Truran, J. W., Primas, F., Collier, J., \& Burles, S. 2005, \apj, 627, 238

\bibitem[]{646}Cuntz, M., Rammacher, W., \& Ulmschneider, P. 1994, \apj, 432, 690

\bibitem[]{648}Cuntz, M., Rammacher, W., \& Musielak, Z. E. 2007, \apj, 657, L57

\bibitem[]{650}Dravins, D., Linde, P., Ayres, T. R., Linsky, J. L.,
 Monsignori-Fossi, B., Simon, T., \& Wallinder, F. 1993, \apj, 403, 412 

\bibitem[]{653}Dupree, A. K., \& Smith, G. H. 1995, \aj, 110, 405

\bibitem[]{655}Dupree, A. K., Hartmann, L., \& Avrett, E. H. 1984, \apj,
  281, L37

\bibitem[]{658}Dupree, A. K., Baliunas, S. L., Guinan, E. F., Hartmann,
L., Nassiopoulos, G. E., \& Sonneborn, G. 1987, \apj, 317, L85 

\bibitem[]{661}Dupree, A. K., Harper, G. M., Hartmann, L., Jordan, C., Rodgers,
 A. W., \& Smith, G. H. 1990a, \apj, 361, L9

\bibitem[]{664}Dupree, A. K., Hartmann, L., \& Smith, G. H. 1990b, \apj,
  353, 623

\bibitem[]{667}Dupree, A. K., Sasselov, D. D., \& Lester, J. B. 1992, \apj, 387, L85

\bibitem[]{669}Dupree, A. K., Hartmann, L., Smith, G. H., Rodgers, A. W.,
 Roberts, W. H., \& Zucker, D. B. 1994, \apj, 421, 542

\bibitem[]{672}Dupree, A. K., Whitney, B. A., \& Pasquini, L. 1999, \apj, 520,
 751

\bibitem[]{675}Fossum, A., \& Carlsson, M. 2006, \apj, 646, 579

\bibitem[]{677}Fulbright, J. P., 2000, \aj, 120, 1841

\bibitem[]{679}Gondoin, P. 1999, \aap, 352, 217

\bibitem[]{681}Gondoin, P. 2005, \aap, 444, 531

\bibitem[]{}Gonzalez, G., \& Wallerstein, G. 1998, \aj, 116, 765

\bibitem[]{683}Gratton, R. G., Pilachowski, C. A., \& Sneden, C. 1984, \aap, 132,
  11

\bibitem[]{686}Hanson, R. B., Sneden, C., Kraft, R. P., \& Fulbright, J.  
 1998, AJ, 116, 1286

\bibitem[]{}Hartmann, L., Dupree, A. K., \& Raymond, J. C. 1982, \apj,
252, 214

\bibitem[]{}Hartmann, L., Jordan, C., Brown, A., \& Dupree,
A. K. 1985, \apj, 296, 576

\bibitem[]{}Janes, K. A., \& Heasley, J. N. 1988, \aj, 95, 762

\bibitem[]{}Janes, K. A., \& Smith, G. H. 1984, AJ, 89, 487


\bibitem[]{}Johnson, H. L., \& Sandage, A. R. 1955, \apj, 121, 616

\bibitem[]{}Johnson, H. L., Iriarte, B., Mitchell, R. I., \&
Wisniewskj, W. Z. 1966, Comm. Lunar Planet. Lab., 4, 99


\bibitem[]{689}Johnson, J. A. 2002, ApJS, 139, 219

\bibitem[]{691}Judge, P. G., \& Stencel, R. E. 1991, \apj, 371, 357

\bibitem[]{}Judge, P. G., Carlsson, M., \& Stein, R. F. 1983, \apj, 597, 1158

\bibitem[]{693}Kemp, S. N., \& Bates, B. 1995, \aaps, 112, 513

\bibitem[]{695} Kurucz, R. L. 2005, Mem. Soc. Astron. Ital. Supp., 8, 14

\bibitem[]{697}Lobel, A., \& Dupree, A. K. 2000, \apj, 545, 454

\bibitem[]{699}Lyons, M. A., Kemp, S. N., Bates, B., \& Shaw, C. R. 1996, \mnras,
  280, 835

\bibitem[]{702}Mallia, E. A., \& Pagel, B. E. J. 1978, \mnras, 184, P55

\bibitem[]{704}Mauas, P. J. D., Cacciari, C., \& Pasquini, L. 2006, \aap, 454,
  609

\bibitem[]{}McWilliam, A., Preston, G. W., Sneden, C., \& Shectman,
S. 1995a, AJ, 109, 2736

\bibitem[]{707}McWilliam, A., Preston, G. W., Sneden, C., \& Searle, L. 1995b, \aj,
  109, 2757
\bibitem[]{}Murray, C. A., \& Clements, E. D. 1968, Royal Obs. Bull.,
No. 139, 309

\bibitem[]{711}Peterson, R. C. 1981, \apj, 248, L31

\bibitem[]{713}Peterson, R. C., \& Schrijver, C. J. 1997, \apj, 480, L47

\bibitem[]{715}Pilachowski, C., Sneden, C., \& Kraft, R., 1996, \aj, 111, 1689

\bibitem[]{717}Pizzolato, N., Maggio, A., \& Sciortino, S. 2000, \aap, 361, 614

\bibitem[]{}Ridgway, S. J. Joyce, R. R., White, N. M., \& Wing,
R. F. 1980, \apj, 235,126

\bibitem[]{719}Rutten, R. G. M., Schrijver, C. J., Lemmens, A. F. P., \& Zwaan, C.
 1991, \aap, 252, 203

\bibitem[]{724}Sandage, A. 1970, \apj, 162, 841

\bibitem[]{}Sanders, W. L. 1977, \aaps, 27, 89

\bibitem[]{722}Schrijver, C. J. 1987, \aap, 172, 111

\bibitem[]{726}Seaton, M. J. 1979, \mnras, 187, 73p

\bibitem[]{728}Shetrone, M. D. 1994, \pasp, 106, 161

\bibitem[]{730}Shetrone, M. D., \& Keane, M. J. 2000, \aj, 119, 840

\bibitem[]{732}Simon, T., \& Drake, S. A. 1989, \apj, 346, 303

\bibitem[]{734}Simon, T., \& Landsman, W. 1991, \apj, 380, 200

\bibitem[]{}Simon, T., Linsky, J. L., \& Stencel, R. E. 1982, \apj,
257, 225

\bibitem[]{736}Smith, G. H., \& Churchill, C. W. 1998, \mnras, 297, 388

\bibitem[]{738}Smith, G. H., \& Dupree, A. K. 1988, \aj, 95, 1547

\bibitem[]{740}Smith, G. H., \& Dupree, A. K. 1998, \aj, 116, 931

\bibitem[]{}Smith, G. H., \& Janes, K. A. 1988 in {\it A Decade of UV
Astronomy with IUE}, ed. E. J. Rolfe (ESA SP-281, Vol. 2), 193

\bibitem[]{742}Smith, G. H., Dupree, A. K., \& Churchill, C. W. 1992, \aj, 
  104, 2005

\bibitem[]{745}Smith, G. H., Dupree, A. K., \& Strader, J. 2004, \pasp, 116, 819

\bibitem[]{747}Soderblom, D. R. 1985, \aj, 90, 2103

\bibitem[]{749}Soderblom, D. R., Duncan, D. K., \& Johnson, D. R. H. 1991, \apj, 
 375, 722

\bibitem[]{}Steiman-Cameron, T. Y., Johnson, H. R., \& Honeycutt,
R. F. 1985, \apj, 291, L51

\bibitem[]{752}Stencel, R., E., \& Mullan, D. J. 1980, \apj, 240, 718

\bibitem[]{}Stencel, R., E., Mullan, D. J., Linsky, J. L., Basri,
G. S., \& Worden, S. P. 1980, \apjs, 44, 383

\bibitem[]{754}Stone, R. P. S. 1983, \pasp, 95, 27

\bibitem[]{756}Strassmeier, K. G., Handler, G., Paunzen, E., \& Rauth, M. 1994,
 \aap, 281, 855

\bibitem[]{}Taylor, B. J.  2007, \aj, 133, 370


\bibitem[]{759}Wedemeyer-B\"ohm, S., Steiner, O., Bruls, J., \&
Rammacher, W. 2007 in  {\it The Physics of Chromospheric Plasmas}, ASP
Conf. Ser. 368, eds. P. Heinzel, I. Dorotovi\v c, \& R. J. Rutten, 93
(astro-ph/0612627)

\bibitem[]{763}Wilson, O. C., \& Bappu, M. K. V.  1957, \apj, 125, 661

\end{thebibliography}
